\documentclass[a4paper,11pt]{article}
\usepackage{graphicx}
\usepackage[space]{grffile}
\usepackage{latexsym}
\usepackage{textcomp}
\usepackage{longtable}
\usepackage{tabulary}
\usepackage{booktabs,array,multirow}
\usepackage{amsfonts,amsmath,amssymb}
\usepackage{natbib}
\usepackage{float}
\usepackage{url}
\usepackage{hyperref}
\hypersetup{colorlinks=false,pdfborder={0 0 0}}
\usepackage{etoolbox}
\makeatletter
\usepackage{authblk}
\usepackage[utf8]{inputenc}
\usepackage[english]{babel}
\usepackage[top=1in, bottom=1in, left=1in, right=1in]{geometry}
\title{Classification of Middle Tropospheric Systems over the Arabian Sea and Western India}
\date{}
\author[1,2]{Pradeep Kushwaha \thanks{Corresponding author: Pradeep Kushwaha, pkushwaha@gmail.com}}
\author[1,2]{Jai Sukhatme}
\author[1,2]{Ravi S. Nanjundiah}
\affil[1]{Centre for Atmospheric and Oceanic Sciences, Indian Institute of Science, Bangalore, 560012, India}
\affil[2]{Divecha Centre for Climate Change, Indian Institute of Science, Bangalore, 560012, India}

\begin{document}
\maketitle

\begin{abstract}
The formation of Middle Tropospheric Cyclones (MTCs) that are responsible for a large portion of annual precipitation and extreme rainfall events over western India is studied using an unsupervised machine learning algorithm and cyclone tracking. Both approaches reveal four dominant weather patterns that lead to the genesis of these systems; specifically, re-intensification of westward moving synoptic systems from Bay of Bengal (Type 1, 51\%), {\it in-situ} formation with a coexisting cyclonic system over the Bay of Bengal that precedes (Type 2a, 31\%) or follows (Type 2b, 10\%) genesis in the Arabian Sea, and finally {\it in-situ} genesis within a northwestward propagating cyclonic anomaly from the south Bay of Bengal (Type 2c, 8\%). Thus, a large fraction of rainy middle tropospheric synoptic systems in this region form in association with cyclonic activity in the Bay of Bengal. The four variants identified also show a marked dependence on large-scale environmental features with Type 1 and Type 2a formation primarily occurring in phases 4 and 5, and Type 2b and Type 2c in phases 3 and 4 of the Boreal Summer Intraseasonal Oscillation.
Further, while {\it in-situ} formation with a Bay of Bengal cyclonic anomaly (Type 2a and 2b) mostly occurs in June, downstream development is more likely in the core of the monsoon season. Out of all categories, Type 2a is associated with the highest rain rate (60 mm/day) and points towards the dynamical interaction between a low pressure system over the Bay of Bengal and the development of MTCs over western India and the northeast Arabian Sea. This classification, identification of precursors, connection with cyclonic activity over the Bay of Bengal and dependence on large-scale environment provides an avenue for better understanding and prediction of rain-bearing MTCs over western India. 
\end{abstract}

\section{Introduction}
Moist synoptic weather systems are an essential component of the Indian summer monsoon \citep{mooley1973,sikka1980some, mooley1987characteristics, krishnamurthy2010composite,patwardhan2020synoptic}.
In particular, monsoon low-pressure systems (called monsoon depressions, if intense) are known to contribute significantly to east and central Indian rainfall \citep{yoon2005water,hunt2016spatiotemporal}. These systems show vorticity and wind maxima in the lower troposphere with closed isobars around the center, usually have a vertically upright structure and primarily move west or northwest from the Bay of Bengal to the Indian landmass \citep{godbole1977,krishnamurthy2010composite,boos2015adiabatic}. The genesis of these low-pressure systems (LPSs) has received enormous attention in the last few decades, and various mechanisms for their formation have been proposed \citep{krishnamurti1977downstream,shukla1978,goswami1980role,AdamesMing,meera2019downstream,diaz2019barotropic,diaz2019monsoon}; further, recent work has brought forth the role of Intraseasonal Oscillations (ISOs) in creating a favorable environment for the genesis of these systems \citep{karmakar2021influence,deoras2021four}. 
In fact, during the summer monsoon, LPSs over the Bay of Bengal have been classified as being born {\it in-situ} or associated with downstream development of South China Sea disturbances \citep{krishnamurti1977downstream,chen1999interannual,meera2019downstream}. 

\noindent In contrast, western India receives rainfall from so-called Middle Tropospheric Cyclones (MTCs), which show maximum intensity in the middle troposphere, have a vertical tilt, and a relatively weak signature near the surface \citep{carr1977mid,francis2006intense,choudhury2018phenomenological,ksn}; notably, LPSs are almost absent during primary monsoon months over this region \citep{HurleyBoos,ksn}. Since a significant amount of rainfall (about 50\%) in western India is accounted by the topographic uplift via the Western Ghats mountains along the west coast of India 
\citep{miller1968iioe,wu1999numerical,tawde2015investigation,zhang2018numerical}, the influence of synoptic systems on rainfall in this region has been somewhat overshadowed. 
Indeed, the west coast of Maharashtra in western India and adjoining the Arabian Sea generally receives moderate-intensity orographic rain throughout the primary monsoon months \citep{miller1968iioe,ramage1971monsoon,kumar2017vertical}. However, it has been observed recently that almost every year, parts of the state (especially Mumbai city and Konkan) face flooding due to cyclonic vortices or MTCs \citep{shyamala2006impact,francis2006intense,kumar2008analysis,choudhury2018phenomenological,ray2019recent}. In fact, the state of Gujarat and the west coast of Maharashtra are regions that receive more than 70\% of their annual rainfall from heavy rainfall events with the intensity of rain events reaching 80 mm per day \citep{kumar2014role,vuruputur2018tropical}.

\noindent Historically, the first MTC over the northeast Arabian Sea was detected, and its impact on western India was realized during the International Ocean Expedition (IIOE). Specifically, during 28 June-10 July 1963, a synoptic system was observed over the northeast Arabian Sea which remained quasi-stationary near the western coast of India. It produced heavy rain from 2-10 July 1963 as recorded by several stations along the west coast of India and Gujarat \citep{miller1968iioe}. While limited data before the formation of this MTC made the cause of genesis challenging to identify, a few essential aspects observed were --- its quasi-stationary nature off the coast of Mumbai for about 12 days and the precedence and coexistence of a LPS over the Bay of Bengal throughout its life cycle. In fact, an enhancement of east-west cyclonic zonal shear over western India between 700-500 hPa with the northwest motion of the Bay of Bengal system was noted on 27 June, following which a MTC was detected on 28 June over the Konkan coast and the northeast Arabian Sea. After the formation of this MTC, an east-west trough stabilized in the middle troposphere and extended from the Arabian Sea to the Bay of Bengal. At the same time, heavy rain started along the west coast of India, reaching up to $22^{\circ}$N. Comparing the surface and 500 hPa vorticity, divergence, and wind fields during this MTC event, \cite{miller1968iioe} concluded that there was hardly any change in the direction, intensity, and convergence of near-surface winds during the transition from light to heavy rainfall. Therefore, the formation of the MTC and enhanced middle troposphere cyclonic shear were the only significant changes that resulted in the extreme rainfall event. In fact, \cite{miller1968iioe} found a correlation of 0.95 between the location of extreme rainfall events over western India and the axis of the middle tropospheric zonal trough. This detailed case study was the first to clearly point to the vital role of MTCs and the middle tropospheric zonal trough in an extreme rainfall event over western India.

\noindent In addition to this case study, \cite{miller1968iioe} analyzed conditions during three additional instances of Arabian Sea MTCs and confirmed that in all the cases, heavy rainfall events occurred with these middle tropospheric systems. In all cases, again, the coexistence of a warm core monsoon low-pressure system in the Bay of Bengal was a persistent phenomenon. Further analysis of lead-lag times series of relative vorticity suggested that the cyclonic anomaly in the Bay of Bengal preceded the extreme rain events and vorticity enhancement over the Arabian Sea and western India \citep{miller1968iioe, ramage1971monsoon}. Along similar lines, \cite{choudhury2018phenomenological} examined 20 heavy precipitating MTCs over western India from the India Meteorological Department (IMD) data and found that 90\% of them formed in the presence of westward-moving LPSs over the Bay of Bengal. They also noted that during the formation of most of the MTCs, the 30--60 day ISO modes were active over western India.

\noindent Regarding the formation of middle tropospheric systems, steady-state monsoon models have been used to argue that the heat low to the west and north of the Arabian Sea exports vorticity at middle levels, which in turn triggers the formation of Arabian Sea MTCs \citep{ramage1966summer}; in fact, the strength of this heat low is known to be partly connected with cyclonic activity over the Bay of Bengal through the subsidence warming of northwest India \citep{ramage1966summer,ramage1971monsoon}. The local formation of synoptic systems in this region is complicated by the presence of deserts to the west and north and narrow western Ghat mountains to the east \citep{krishnamurti1981onset,varikoden2019contrasting}. Despite this, fundamental dynamical instability investigations have probed the {\it in-situ} formation of MTCs \citep{goswami1980role,brode1978mechanism,mak1975monsoonal,mak1983moist,mak1982instability}; however, complexities in identifying proper mean state for instability analysis in an evolving monsoon flow have plagued these efforts \citep{carr1977mid}. Apart from attempts at identifying the {\it in-situ} formation of Arabian Sea systems, surveys of satellite images showed MTCs over the Arabian Sea to be remnants of monsoon lows moving westward from the Bay of Bengal \citep{carr1977mid}. 
Moreover, detailed manual tracking of MTCs over South Asia using modern reanalysis data spanning sixteen years clearly showed that many MTCs originate as monsoon lows over the Bay of Bengal and later behave like MTCs once they reach the Arabian Sea \citep{ksn}. 
Thus, in addition to {\it in situ} genesis, downstream development of LPSs from the Bay of Bengal is a potential route for the formation of MTCs over the northeast Arabian Sea and western India.  

\noindent In terms of meteorological and societal impact, not only does western India witness some of the world's heaviest rainfall events and associated floods \citep{mapes2011heaviest,vuruputur2018tropical,choudhury2018phenomenological}, it has also shown an increasing trend in both annual precipitation and in the frequency of extreme rain events in the past few decades \citep{rajeevan2008analysis,pattanaik2010variability,vinnarasi2016changing,roxy2017threefold}. As cyclonic systems over western India and the northeast Arabian Sea are confined to the middle troposphere during most of their life cycle, they remain unrecognized by traditional surface weather charts and are usually underestimated by tracking algorithms that use lower troposphere fields as identifying markers \citep{miller1968iioe,ramage1971monsoon,ksn}.
Thus, despite the crucial role of synoptic middle troposphere systems in extreme rain events and annual rainfall over western India, no comprehensive work is available on their classification and formation mechanisms. 
For example, it is not clear how many of these systems form locally and what fraction results from downstream development of LPSs from the Bay of Bengal. Or, for that matter, whether specific categories of MTCs are especially prone to producing heavy rainfall. In all, there is a pressing need to classify MTCs and understand their precursors using a large sample size afforded by modern reanalysis data. This will allow proper investigation of local instability mechanisms for specific classes and more broadly open a doorway for the potential predictability of extreme rainfall events over western India. 

\noindent Here, after a description of the data used and the details of the methodology employed for analysis (Sections 2 and 3), we begin exploring the routes of formation of Arabian Sea MTCs by analyzing characteristics of rainy days in western India (Section 4) and whether there are distinct weather patterns associated with a bulk of the precipitating events. To probe this in an unbiased manner, we adopt an unsupervised machine learning ($k$-means clustering) technique to extract possible clusters of weather patterns of rainy days (Section 5). Further, we use an independent method to identify and track middle tropospheric systems and then classify them as per their formation mechanism and compare the results with the $k$-means methodology. Note that while cyclonic systems in the northeast Arabian Sea and western India show vorticity maximum in the middle troposphere, however, at times, they also descend to the surface and behave like monsoon lows \citep{ksn}.
Thus, irrespective of the occasional low-level vorticity maximum, we consider the Arabian Sea and western Indian systems in a single class of "Arabian Sea MTCs," which includes all possible rain-bearing cyclonic systems in this region. In particular, cyclone tracking is done using mid-level fields that captures all synoptic systems in this region (Section 6). Tracking allows us to quantify what fraction of systems are born locally and how many are due to the downstream development (or are possibly remnants) of systems that formed over the Bay of Bengal or east India. Further, cyclone tracking also allows for an examination of track density, motion vectors, and the genesis density of these systems. Moreover, given the apparent link between cyclonic activity over the Bay of Bengal and Arabian Sea systems, we systematically explore this propensity of coexistence in terms of how many Arabian Sea MTCs form in the presence of a cyclonic system over the Bay of Bengal. Precipitation characteristics and monthly frequency within a season are also explored in Section 6, while the modulation by ISOs of each category is studied in Section 7. Finally, our conclusions are presented in Section 8. We note the dynamical interaction between cyclonic activity over the Bay of Bengal and the formation of MTCs over the Arabian Sea is explored in a companion paper.


\section{Data}

\noindent The main product used in this work is the ECMWF ERA-5 fifth-generation atmospheric reanalysis data set \citep{hersbach2020era5} which is generated using 41r2 of the Integrated Forecast System (IFS) model. IFS system utilizes a four-dimensional variational data assimilation scheme and takes advantage of 137 vertical levels and a horizontal resolution of $0.28125^{\circ}$ ($~31$ km, or TL639 triangular truncation). The data is stored at every hour of model integration. This study utilizes six-hourly winds, vorticity, divergence, temperature, and moisture fields on pressure levels from $1000-100$ hPa and native horizontal resolution. However, we use interpolated data on 1.5 degrees latitude-longitude gird for synoptic charts. Apart from a high spatial and temporal resolution, ERA-5 has several important updates to its predecessor ERA-I, which was terminated in 2019. 
These include the use of ozone, satellite radiance, aircraft, and surface pressure data in the assimilation scheme. One of the important change in ERA-5 is the use of an all-sky approach instead of the clear sky approach used in ERA-I, thus providing additional information about precipitation and cloud distribution. These updates, along with others, have resulted in more consistent sea-surface temperature and sea-ice compared to ERA-I \citep{hersbach2020era5}.

\noindent The latest high-resolution reanalysis products, such as ERA-5, that ingest large amounts of satellite and ground-based observations, have been successfully employed in detecting and tracking Indian monsoon LPSs and understanding their contribution to monsoon rainfall \citep{HurleyBoos,Hunt,boos2015adiabatic,hunt2019relationship}.
In fact, modern reanalysis data has allowed for reliable investigations of different structural aspects of monsoon lows, such as their vertical thermal and dynamical features and motion characteristics \citep{Hunt,hunt2019relationship,sorland2015dynamic}. Of course, despite significant improvement in reanalysis products, there are still several caveats to be kept in mind; these include unresolved low-level structure and an exaggerated cold-core in the lower troposphere \citep{manning2007evolution,wood201440}, and at times, unrealistic thermal structures, especially in regions of the globe that have a paucity of observations \citep{janiga}. Further, the intensity of systems, their temporal coherence, and structure is likely to differ among different reanalysis products \citep{hodges2003comparison}. Nevertheless, in light of several improvements, we believe that ERA-5, the highest resolution reanalysis, can be used for the present goals \citep{mahto2019does,nogueira2020inter,yeasmin2021detection,bian2021well}. It is important to note that, as a precaution, we use the reanalysis in conjunction with satellite-based products for robustness and to avoid false detection of rain-bearing systems. 

\noindent To understand the role of large-scale environmental conditions in Arabian Sea MTCs formation, we utilize intraseasonal Boreal Summer Intraseasonal Oscillation (BSISO) indices which are available at \url{http://iprc.soest.hawaii.edu/users/kazuyosh/Bimodal_ ISO.html} [Accessed May 2021]. 
This includes the first two normalized principal components ($PC1$ and $PC2$) with their respective magnitude ($\sqrt{PC1^{2}+PC2{^2}}$) and phase \citep{kikuchi2010formation,kikuchi2012bimodal,kikuchi2020extension}, that explain about 25\% variance of the Outgoing Longwave Radiation (OLR) and wind fields at intraseasonal scales \citep{wheeler2004all}. This data is available at a daily temporal resolution from 1979-2020. Each phase of this intraseasonal oscillation (ISO) corresponds to a specific location of enhanced moist convection over South Asia, and its amplitude represents the strength of this enhanced convection. In particular, phase 2 to phase 3 of BSISO corresponds to enhanced convection over the southern Arabian Sea and the southern Bay of Bengal. Phases 4 \& 5 are when the Indian subcontinent goes through a wet spell and convection shifts to the northern Bay of Bengal and northern Arabian Sea. During Phases 6 through 8, convection weakens, representing the dry or break phase over the Indian subcontinent. Several authors have suggested that the BSISO phase and amplitude modulate various meteorological phenomena including the onset of monsoon, formation of cyclones, and monsoon lows \citep{wheeler2004all,karmakar2021influence,hunt2021modes,deoras2021four}. Thus, it is worth exploring whether the BSISO also influences the formation of synoptic systems in the Arabian Sea and western India. In general, two components usually describe ISO modes --- the eastward-moving Madden-Julian Oscillation, which dominates in boreal winter (December-May), and the northward propagating BSISO, which is prominent during the boreal summer (May-October). As we are dealing with the summer monsoon, we mainly use the BSISO index to represent the strength and phase of intraseasonal activity. 

\noindent We also use Tropical Rainfall Measuring Mission (TRMM) data for rainfall estimates \citep{huffman2007trmm, huffman2010trmm}. In particular, the TRMM Multi-satellite Precipitation Analysis (TMPA)-3B42 Version 7 product with a horizontal resolution of $0.25^{\circ}\times 0.25^{\circ}$ is used from 1998-2019. The improvements in Version 7 of this product are described in \citep{prakash2013comparison}. This data is utilized to distinguish rainy cyclonic systems from dry ones and to understand the quantum of precipitation in different classes of systems. Finally, OLR data at $2.5^{\circ}$ spatial and daily temporal resolution from the National Oceanic and Atmospheric Administration \citep{liebmann1996description} is used as a proxy for moist convection.


\section{Methodology}

\subsection{Identification of weather patterns}

\noindent The identification of distinct weather patterns associated with rainfall over western India is achieved by clustering of daily 600 hPa geopotential height anomaly maps by an unsupervised machine learning method known as $k$-means \citep{hartigan1979ak}. Prior to applying the method, only moderate or higher intensity rain days are retained by removing dry and light rain days using the threshold $P_{WI}>7.6$ mm/day, as defined by the Indian Meteorological Department; where $P_{WI}$ is the mean TRMM rainfall over western India bounded by $68^{\circ}$E-$72^{\circ}$E and $15^{\circ}$N-$22^{\circ}$N; this region of averaging is similar to that of \citep{choudhury2018phenomenological} with a slight reduction of eastern boundary and increase to the south to avoid the effect of topographic rainfall since we are primarily interested in precipitation from synoptic-scale systems.

\noindent In the recent past, $k$-means clustering has proved to be a useful tool in a variety of meteorological applications \citep{awan2015identification,jiang2016global,pope2009regimes,clark2018rainfall}. Broadly, clustering is an approach wherein similar data vectors (daily 600 hPa height anomaly maps in this study) are placed into unique groups or clusters. In $k$-means, a set of $x$ observations, [$X = X_{1},X_{2},X_{3}....X_{i}...X_{x}$]; which are real vectors, each having $N$ dimensions are sorted into $k$ clusters, [$K = {K_{1},K_{2},K_{3},.... K_{i}....K_{k}}]$. Here, we use daily geopotential height anomalies of 600 hPa pressure surface from the domain $0-30^{\circ}$N and $60-100^{\circ}$E as the set of vector arrays, with each observation having dimension $N = m\times n$; where $m$ and $n$ are the number of latitude and longitude points in the domain. 
As similar data vectors will have small Euclidean distances, they will be placed into the same clusters. In the context of this work, this amounts to placing height anomaly patterns with similar spatial distribution and intensity into the same cluster. Note that the domain on which we perform $k$-means includes the dominant regional features of monsoon: specifically, the monsoon trough, heat low to the north, the Bay of Bengal to the east, and low-level monsoon jet to the south. 

\noindent Essentially, the $k$-means clustering algorithm involves the following steps:
\begin{enumerate}
    \item Randomly assigns $k$ data points as initial cluster centroids.
    \item Compute the Euclidean distance of all data points from each cluster centroid. 
    \item Assign each data point to a cluster centroid with the minimum Euclidean distance.
    \item Recalculate the cluster centroids and repeat steps II and III until each cluster centroid becomes stable. 
\end{enumerate}

\noindent This method heavily relies on the parameter $k$, i.e., a number of user-defined initial clusters. Therefore, it is crucial to select the appropriate number of clusters $k$; a too-small value of $k$ may result in merging different clusters which may have essential differences thus susceptible to the loss of important weather patterns; too large a value may result in multiple clusters which may be only slightly different in intensity or spatial patterns. 
Here we used two complementary methods to select $k$; the silhouette coefficient \citep{rousseeuw1987silhouettes} and elbow method \citep{bholowalia2014ebk,syakur2018integration}. The mean and individual silhouette coefficient (SC) is calculated for different values of $k$ (2 to 10) for each cluster and is shown in Figures S1 and S2, respectively. In general, higher positive values of SC indicate better clustering. Here, a relatively high value of SC is found from $k=2-4$, and it drops sharply after that. Hence, from SC, we say that $k \in [2,4]$ can be used as an appropriate choice. To further constrain the value of $k$, we then use the elbow method. In this method, the elbow point of normalized error (i.e., Within Cluster Sum of Squares Distances: WSS) curve is taken as the optimal value $k$, i.e., the value of $k$ at the elbow point minimizes the total WSS. The location of the knee point in our analysis is calculated by a widely used objective method called kneedle \citep{satopaa2011finding}, which suggests a knee point at $k=4.27$ (Figures S3 and S4). Apart from object guidance for the selection of $k$, it is appropriate to check for the physical significance of obtained clusters \citep{clark2018rainfall}. Indeed, once the cluster analysis was completed, we assessed that the physical meaning of the obtained patterns was consistent with our knowledge of Arabian Sea systems. With the aid of both the objective methods and the inspection of their physical validity, the value $k=4$ was deemed most appropriate for our analysis.

\noindent We used the python toolbox "kmeans" to implement the $k$-means clustering algorithm. The model is initialized with $k$ random data points as initial centroids. Since convergence occasionally becomes sensitive to initialization, we ran the model for 600 random initial centroids and finally selected one which minimizes the sum of the cluster of squares distances (i.e., inertia). Further, 100 iterations are used for every single run. In general, we find that the WSS decreases for the first 20 or so iterations and then settles down with no further change in centroids. Hence, 100 iterations are sufficient to ensure that the algorithm has converged to a solution.

\subsection{Detection and tracking of systems}

\noindent Cyclone center detection is performed using 600 hPa geopotential height field as Arabian Sea and western Indian systems are known to show maximum intensity in the middle troposphere \citep{ksn}. At each six-hourly time step, a local minimum of geopotential height is detected as a system center. After that, only strong system centers are retained for the formation of tracks by applying the threshold $\xi_{m}>1\times10^{-5}s^{-1}$, where $\xi_{m}$ is the mean vorticity around $4^{\circ}$ area around the detected centers \citep{ksn}. The formation of tracks from the above cyclone centers is achieved by a standard nearest neighborhood method. In particular, we use the first guess method of \cite{hanley2012objective}, initially used for extratropical system tracking \citep{wernli2006surface}. Variants of this method have also been employed in the tropics for tracking of monsoon lows \citep{praveen2015relationship}. In this method the first-guess location of cyclone candidate is a reduced linear continuation of the track in geographical longitude-latitude: $P^{*}(t_{(n+1)})$ = $P(t_{n}) + q\times[P(t_{n}) - P(t_{n-1})]$, where $P$ is a latitude-longitude coordinate pair, $t_{n}$ is the $n$th time step and $q$ is a factor which weights the forward speed of the cyclone. A cyclone center located at $P$ at time $t_{t+1}$ regarded as the new position if it minimizes the distance among all cyclones present at $t_{n+1}$ within the radius $D_{0}$. 
For robustness, we also employed vorticity threshold wherein the difference between predicted vorticity, $\xi^{*}(t_{(n+1)})$ = $\xi(t_{n}) + r \times [\xi(t_{n}) - \xi(t_{n-1})]$ and the vorticity of the nearest neighbour candidate must be less then a threshold value of $\xi^{*}(t_{n+1})$ - $\xi_{n+1}<5 \times 10^{-5}s^{-1}$. Here, we use tuning parameters $q=1$ and $r=0.75$ for optimal tracking \citep{hanley2012objective, praveen2015relationship}. If no valid candidates are found for consecutive 12 hours, then tracks are terminated. Further, only tracks that last for more than three days are retained. Results are validated against manual maps of extended MTC data sets \citep{ksn} (Figure S5), and those of monsoon lows \citep{Hunt,praveen2015relationship}. Moreover, we also check individual tracks for selected cases against the daily motion of cyclones using six-hourly map animations. For clarity, a schematic of the tracking procedure is shown in Figure S6. 

\section{Results}
\section{Large Scale Conditions During Western India Rainfall}

\begin{figure*}
\centering
\includegraphics[trim=0 0 0 0, clip,height = 1\textwidth,width = 0.75\textwidth, angle =90, clip]{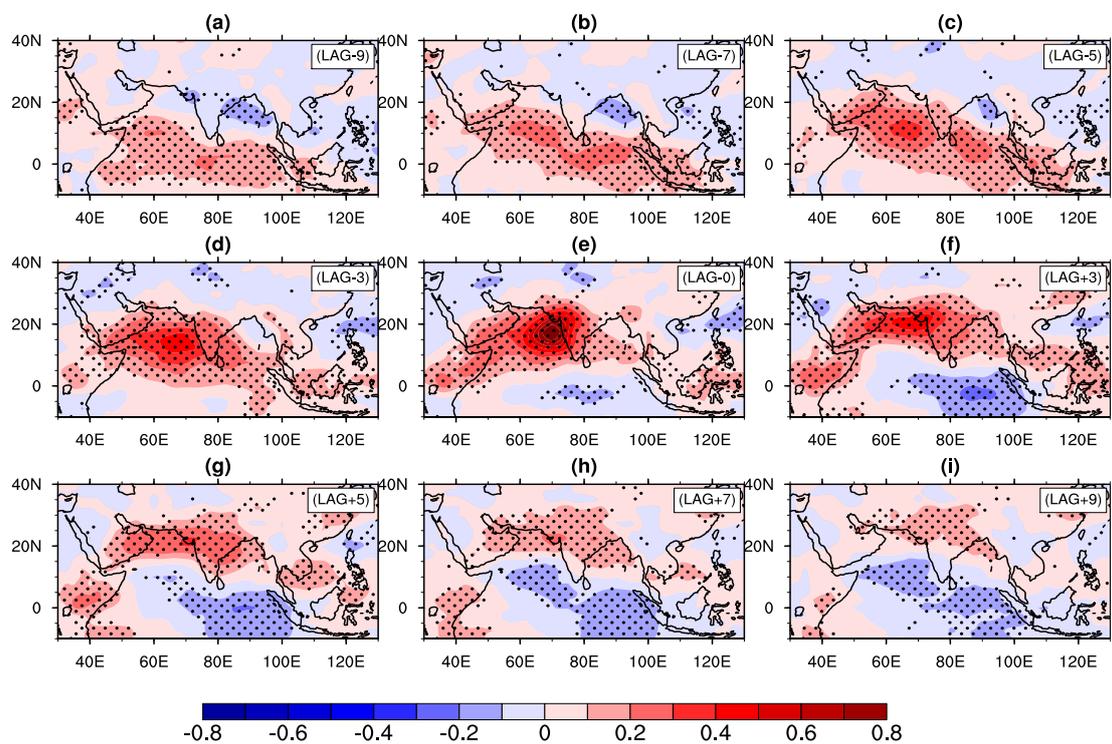}
\caption{Lag correlation of OLR anomaly over the shown domain with the Arabian Sea ($15-22^{\circ}$N to $68-72^{\circ}$E) mean OLR anomaly time series from June to September, 1980-2019. Shading denotes the regions which are significant at 99\% confidence.}
\label{fig:FIG1}
\end{figure*}

\noindent Before diving into the classification and formation mechanisms of synoptic systems, it is worth exploring the prevalent large-scale conditions during rainfall over western India. The lag-correlation of daily OLR --- a proxy for moist convection --- in the South-Asian sector 
with the mean daily OLR time series over western India ($68-72^{\circ}$E, $15-22^{\circ}$N) for 37 years during June through September is shown in Figure~\ref{fig:FIG1}. Here, Day 0 represents the correlation at zero lag. At $8-10$ days lag (Figure~\ref{fig:FIG1}a), a large-scale east-west elongated belt of positive correlation appears near the equator --- reflecting the presence of Intertropical Convergence Zone (ITCZ) near the equator or Phase 1 of BSISO  \citep{kikuchi2012bimodal,lee2013real}. This region of positive correlation stretches from the western equatorial Indian Ocean up to the South China Sea. From Day $-9$ to 0 (Figure~\ref{fig:FIG1}a-e), northward propagation of correlation patterns are pronounced over the Arabian Sea and Bay of Bengal.
Interestingly, the Arabian Sea branch appears to advance much faster than its Bay of Bengal counterpart. These differences in the nature of propagation result in a northwest to southeast tilt, which has been observed in prior studies \citep{karmakar2020differences}. The northward movement is seen up to Day 0 over both basins; after which the Bay of Bengal branch shows a west-northwest extension from Day $0$ to $3$ (Figure~\ref{fig:FIG1}e-f) that is maybe a reflection of relatively high-frequency variability associated with the westward motion of monsoon lows and Rossby waves \citep{goswami1987mechanism,sobel2000dynamics,karmakar2020differences,karmakar2021influence}. Moreover, all the way through Days 0 to 9 (Figure~\ref{fig:FIG1}e-i), while regions of positive correlation gradually weaken, they are seen to remain quasi-stationary, with one maximum over the northeast Arabian Sea and the other over the east coast of India and the Bay of Bengal, indicating the co-existence of convective activity over both these regions. Since the large-scale correlation pattern concurrently covers both the Arabian Sea and the Bay of Bengal, it suggests favorable conditions for cyclonic systems over both the basins simultaneously. Hence, it is not surprising that prior studies reported the co-existence of synoptic systems over these two regions \citep{ramage1971monsoon,carr1977mid,choudhury2018phenomenological}. Although the correlation of OLR anomalies includes the entire spectrum of temporal variability, the spatio-temporal evolution of these patterns closely resembles those of intraseasonal modes \citep{kikuchi2012bimodal,lee2013real}.   
Essentially, these results suggest that the rainfall over western India and the eastern Arabian Sea is likely modulated by northward propagating large-scale OLR anomalies that are similar to the BSISO \citep{kikuchi2012bimodal}. 

\subsection{Rainy day composites}

\begin{figure*}
\centering
\includegraphics[trim=0 0 0 0, clip,height = 1\textwidth,width = 0.65\textwidth, angle =0, clip]{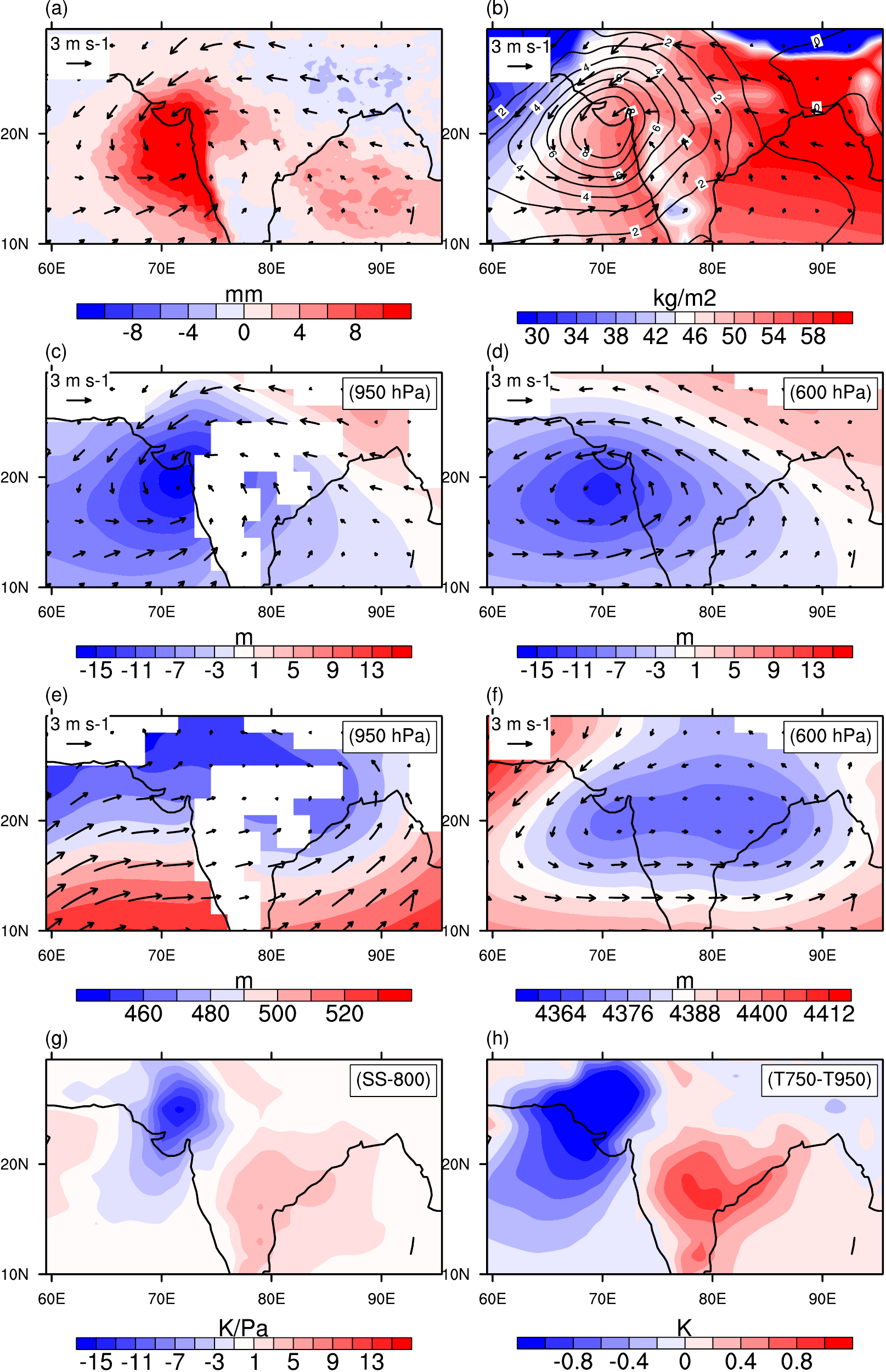}
\caption{Composites of rainy days ($P_{WI}>7.6$ mm/day) (a) composite rain anomaly (b) composite precipitable water anomaly (contours: kg/m2) and June-September climatology (colors); (c) composite geopotential height anomaly at 950 hPa for rainy days; (d) same as (c) but for 600 hPa; (e), (f) composite total geopotential heights at 950 and 600 hPa, respectively; (g) is the static stability parameter at 800 hPa; (h) anomalous strength of inversion ${\Delta}T_{ano} = T_{a750}-T_{a950}$, where $T_{a750}$ and $T_{a950}$ are temperature anomalies at 750 and 950 hPa. heights are in meter and rainfall in mm; the white regions are where topography crosses the geopotential surface of analysis.} 
\label{fig:FIG2}
\end{figure*}

\noindent To understand the dynamical features prevalent during rain events over western India, we now construct composites of meteorological fields on rainy days. 
In particular, composites of the precipitation anomaly, precipitable water anomaly, total and anomalous wind, and geopotential height anomalies during the rainy days over western India at 950 hPa and at 600 hPa are shown in Figure~\ref{fig:FIG2}; anomalies are with respect to daily climatology constructed from 22 years of reanalysis data. 

\noindent During rainy days, precipitation anomalies (Figure~\ref{fig:FIG2}a) are largest along the western coast of India but extend up to central and east India. Similarly, precipitable water (Figure~\ref{fig:FIG2}b) is anomalously high over the northeast Arabian Sea and western India with a positive anomaly up to the Bay of Bengal. The precipitable water anomaly maximum (Figure~\ref{fig:FIG2}b; contours) also coincides with the anomalous circulation, suggesting that anomalous winds might play a role in controlling moisture accumulation. The wind direction over this region is crucial in moisture control, given strong meridional and zonal moisture gradients. Specifically, towards the north-northwest (east), there is a substantial decline (rise) in climatological total column water vapor around $20^{\circ}$N and $70^{\circ}$E (Figure~\ref{fig:FIG2}b; colors). 
Consistent with the anomalous flow, the height anomalies (Figure~\ref{fig:FIG2}c,d) at 950 and 600 hPa depict a widespread low in the middle troposphere over western India and the northeast Arabian Sea. 
In addition, the height anomaly and associated cyclonic circulation over the eastern Arabian Sea and western India (Figure~\ref{fig:FIG2}e,f) appears to be the part of the middle-tropospheric zonally oriented monsoon trough, which extends from Bay of Bengal to the Arabian Sea during the rainy days. 

\noindent The strength of inversion (Figure~\ref{fig:FIG2}h) --- difference of temperature anomaly at 750 and 950 hPa levels ($\Delta T =$   $T_{a750}$-$T_{a950}$) --- weakens during rainy days over the northeast Arabian Sea. This is also reflected in the reduction of the static stability at the 800 hPa level (Figure~\ref{fig:FIG2}g). This is reasonable as the prevailing anomalous northeasterly winds reduce the advection of cold maritime air at low levels, while enhanced easterly winds in the middle troposphere prevent the advection of warm, desert air from the northwest at middle levels, effectively reducing the strength of inversion. Indeed, this is consistent with the noted decrease and increase in strength of inversion in this region during the active and break phases of the Indian monsoon, respectively \citep{narayanan1981detection,dwivedi2021variability}. In fact, the reduction in strength of inversion was also noted by \cite{miller1968iioe} in their original study of a MTC. The weakened low-level inversion facilitates a favorable environment for monsoon rain in western India. Once the inversion is destroyed, water vapor can rise vertically and trigger deep convection. 
In addition, consistent with the observations of \cite{miller1968iioe}, we note that north and south of about $20^{\circ}$N (Figure~\ref{fig:FIG2}d), the mid-level anomalous circulation is easterly and westerly, respectively. 

\noindent Thus, as a whole, when western India experiences significant precipitation, the composites clearly point towards the presence of well-developed middle tropospheric anomalies in circulation as well as the geopotential height, both of which are collocated with the anomalous build-up of precipitable water. Furthermore, the erosion of the usually strong inversion layer over the eastern Arabian Sea is notable and consistent with the occurrence of deep convection. 

\section{Weather Regimes over Western India}
\begin{figure*}
\centering
\includegraphics[trim=0 0 0 0, clip,height = 1\textwidth,width = 0.7\textwidth, angle =90, clip]{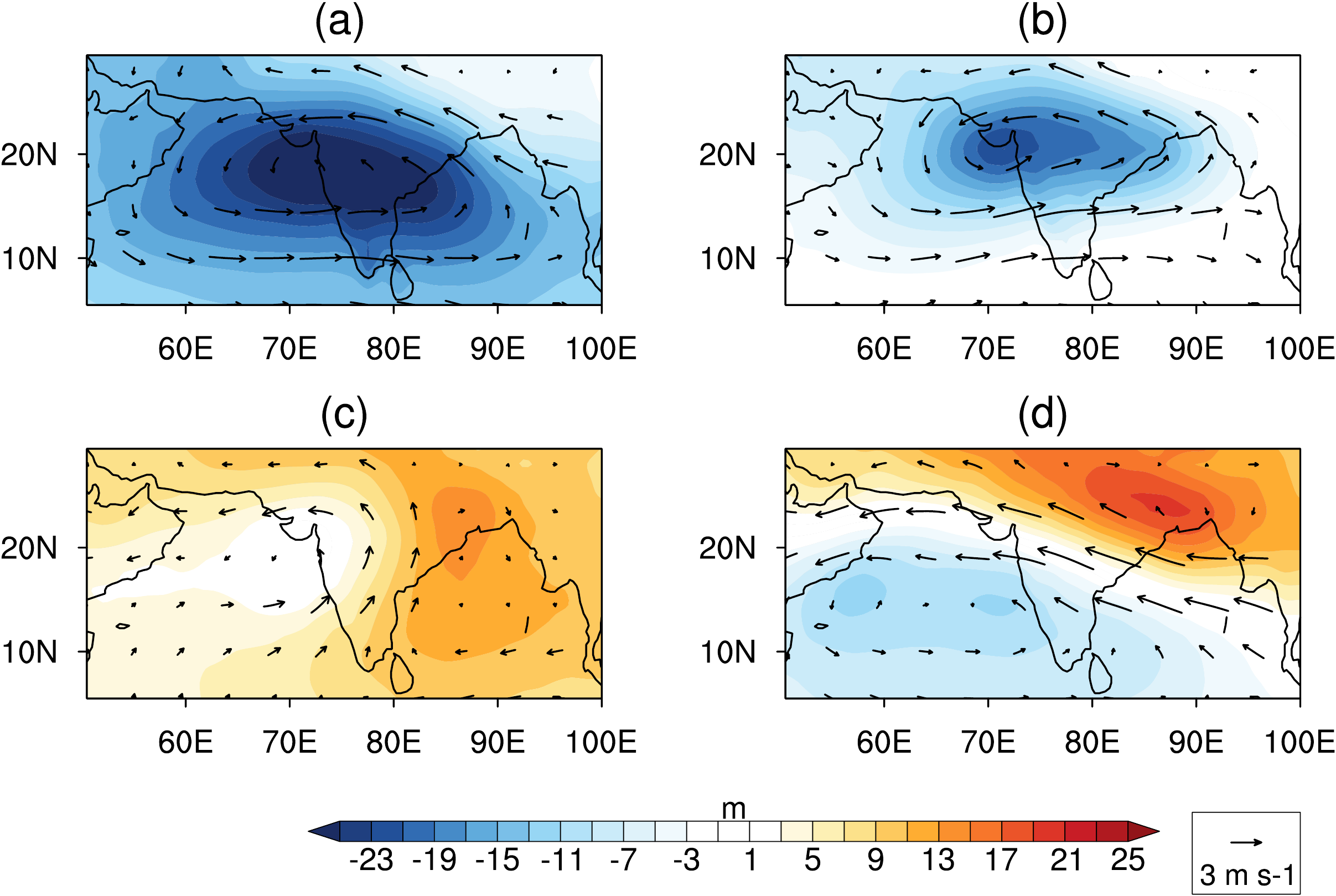}
\caption{Composite height anomaly of 700 hPa surface (m) from $k-$means clustering for Regime 1 (a) to Regime 4 (d). The respective composite winds are shown as arrows, units are m/s.} 
\label{fig:FIG3}
\end{figure*}

\noindent To identify dominant weather regimes or patterns associated with rainy days over western India, and in turn, their precursors, we utilize a $k$-means clustering approach. As discussed, we choose four clusters ($k=4$), and the composite geopotential height and wind anomalies of these four weather patterns as identified by $k$-means are shown in Figure~\ref{fig:FIG3}. As expected, all four dominant weather regimes show anomalous cyclonic circulation and height depression in the middle troposphere over the northeast Arabian Sea and adjoining western Indian region. 
Regime 1 (Figure~\ref{fig:FIG3}a) consists of a sizeable negative height anomaly oriented northwest to the southeast, which covers both the Arabian Sea as well as the Bay of Bengal. This reinforces the notion that Arabian Sea systems exist with cyclonic conditions over the Bay of Bengal \citep{miller1968iioe,ramage1971monsoon}. Regime 2  (Figure~\ref{fig:FIG3}b) is more compact than Regime 1 and consists of a zonally oriented height anomaly that again extends from the Arabian Sea up to the Bay of Bengal. Regime 3  (Figure~\ref{fig:FIG3}c) consists of a relatively weak broad negative height anomaly accompanied with cyclonic circulation over the northeast Arabian Sea, which extends up to the southwest Arabian sea and depicts a northwest to northeast orientation. In this regime, the Bay of Bengal shows a gigantic positive height anomaly and associated anticyclonic circulation. Finally, Regime 4  (Figure~\ref{fig:FIG3}d) shows a zonally elongated negative height anomaly and cyclonic circulation that is closer to the equator and spans a small portion of the southern Bay of Bengal, the southern tip of India, and stretches into the southern Arabian sea. Similar to Regime 3, Regime 4 also shows the signature of a positive height anomaly and anticyclonic circulation over northeast India and the north Bay of Bengal; however, in contrast to Regime 3, here the positive height anomalies are narrow, zonal, and confined to east-central India and foothills of Himalayas with a maximum over West Bengal.
Essentially, the first two regimes indicate cyclonic vorticity and a convectively active environment that ranges from the Arabian Sea \& western Indian up to the Bay of Bengal. In contrast, the latter two regimes suggest the existence of Arabian Sea cyclonic anomalies in isolation which remain relatively weak in the absence of convectively active cyclonic conditions over the Bay of Bengal. 

\begin{figure*}
\centering
\includegraphics[trim=0 0 0 0, clip,height = 1\textwidth,width = 0.6\textwidth, angle =90, clip]{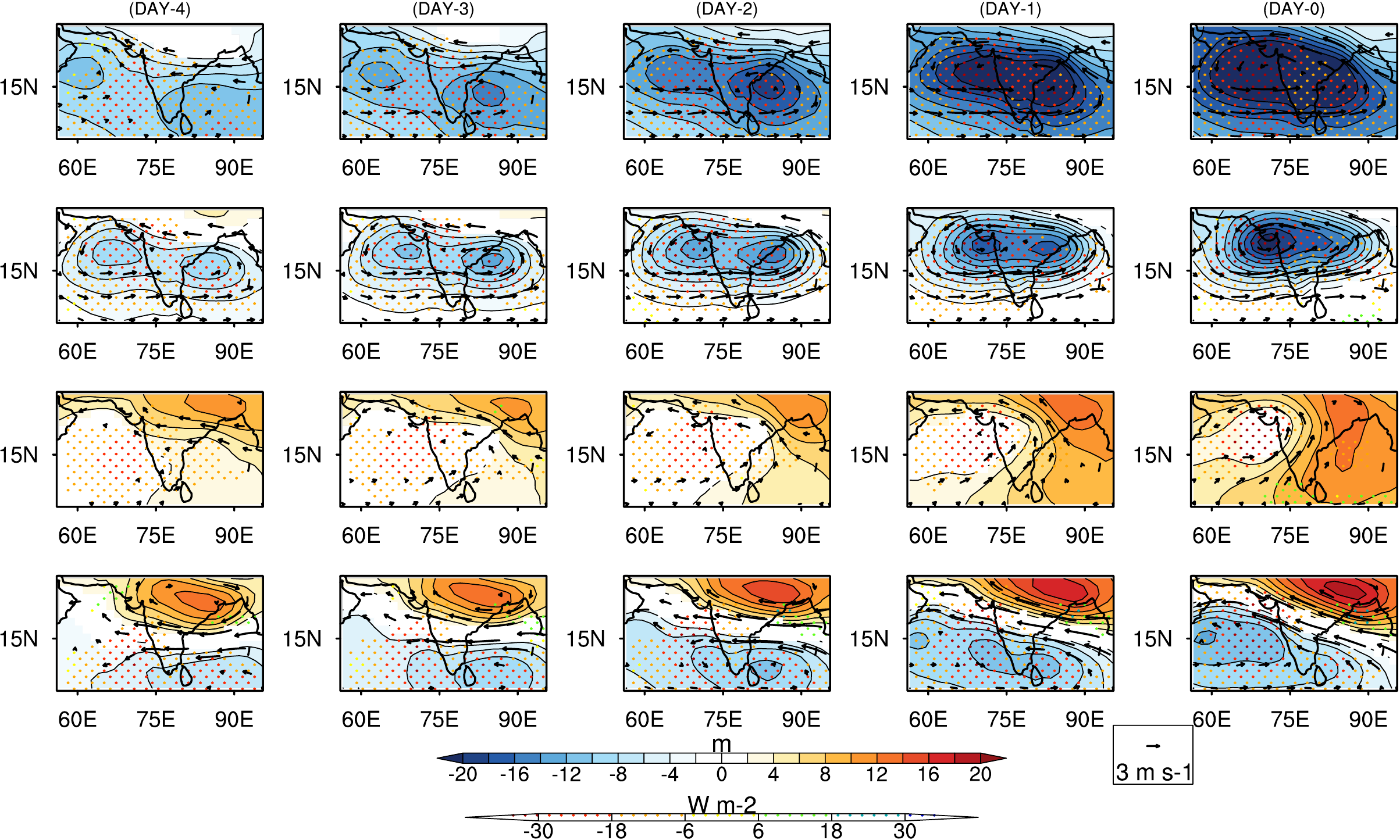}
\caption{Lag composites of the four $k-$means clustering regimes from Day $-4$ to Day 0; row 1 to row 4 represent Regime 1 to 4, respectively. Here, day zero represents the day when the cluster was detected. Color shading represents the composite 600 hPa height anomaly (m), dotted shading represents OLR anomaly, and arrows represent the composite wind anomaly. OLR and height fields are only shown if they are significantly different from climatology at 0.1 significance. Wind vectors are shown if any wind component is significantly different from zero at the 0.1 significance level under two tailed $t$-test. } 
\label{fig:FIG4}
\end{figure*}

\noindent For these four patterns (Regimes 1, 2, 3 \& 4), precursors are now extracted from daily time-lag composites \citep{clark2018rainfall}; these are shown in Figure~\ref{fig:FIG4}. Lag composite of the Regime 1 (row 1; Figure~\ref{fig:FIG4}) suggests a cyclonic circulation and height depression over the Bay of Bengal accompanied by a weak trough-like signature over the Arabian Sea on Day $-4$. Subsequently, from Day $-4$ to $-2$, the Bay of Bengal height anomalies deepens and propagates northwestwards. Following this, from Day $-2$ to Day $-1$, the cyclonic circulation over the northeast Arabian Sea shifts eastwards and merges with the intensifying and northwestwards moving Bay of Bengal anomaly. Essentially, in this regime, the Arabian sea cyclonic conditions appear to be the direct result of westward-moving Bay of Bengal cyclonic anomalies. In essence, westward-moving LPSs in the Bay of Bengal and East India is expected to be a precursor for the formation of this particular class of Arabian Sea systems.

\noindent The lag composite of Regime 2 (row 2; Figure~\ref{fig:FIG4}) shows two dominant centers of action, one over the Arabian Sea and one over the Bay of Bengal from Day $-4$ to Day $-1$. The composite winds and height anomalies of this Regime indicate the presence of twin vortices joined together by an elongated trough line in the middle troposphere with a northwest orientation. This structure is similar to the evolution of the July 1963 MTC wherein, after intensification, the Arabian Sea MTC became joined with the Bay of Bengal LPS by a zonal middle tropospheric trough \citep{miller1968iioe}. The daily evolution of lag-composite shows that initially (Day $-4$) there is a sign of cyclonic systems in both basins, but height anomalies (color shading) deepens first over the Bay of Bengal (Day $-3$), and then the deepening of height anomalies or intensification of the Arabian Sea system follows (Day 0). In fact, a gigantic circulation develops by about Day $-2$ that encircles both systems. To some extent, the structure, orientation, and propagation characteristics of this regime resemble that of phases 4 \& 5 of the BSISO \citep{kikuchi2012bimodal}. 
We anticipate that the co-existence of systems in both basins may have a far-reaching influence on rainfall patterns and intensity by the cooperative intensification through enhanced moisture exchange and an enhanced vorticity-rich environment. In fact, the presence of cyclonic anomaly over both the basins in this regime is in agreement with the notion that most Arabian Sea systems exist in the presence of Bay of Bengal disturbances \citep{carr1977mid,choudhury2018phenomenological,miller1968iioe}.

\noindent The composite winds, height anomaly, and OLR for Regime 3 are shown in Figure~\ref{fig:FIG4}; row 3. Here, weak negative OLR and height anomalies are seen over most Arabian Sea from Day $-4$ to Day $-3$. Then, east India and western regions of the Bay of Bengal also develop a signature of cyclonic circulation around Day $-3$. Hence, as the Arabian Sea system evolves, we note the appearance of cyclonic activity over the west Bay of Bengal. Subsequently, the Arabian Sea height anomaly moves northwards and by Day 0 becomes a concentrated vortex over the northeast Arabian Sea \& western India. In fact, the emergence of this compact vortex causes the cyclonic circulation to withdraw from the Bay, and from Day $-1$ to Day 0, a positive height anomaly appears in the north Bay of Bengal. Thus, when mature, the Arabian Sea and north Bay of Bengal anomalies exhibit an opposite sense of circulation in Regime 3. 

\noindent In Regime 4 (row 4; Figure~\ref{fig:FIG4}), the negative height anomaly in the Arabian Sea system is positioned at a lower latitude compared to Regimes 1,2 \& 3. Here, the north Bay of Bengal and core monsoon zone is characterized by positive height anomalies, anticyclonic circulation, and absence of convection. In contrast to the north Bay of Bengal, a large region of negative OLR and height anomalies exist in the southern Bay of Bengal on Day $-4$. Following this, from Day $-4$ to Day $-3$, the convectively active region of the South Bay of Bengal moves northwestwards into the Arabian Sea (by Day 0). With the movement of this large-scale envelope of negative height anomaly over the Arabian Sea, the genesis of a cyclonic system is observed over the northeast Arabian Sea. Similar to Regime 3, Regime 4 also shows a positive height anomaly over the north Bay of Bengal and Indian monsoon trough region. It is also worth noting that Regimes 3 \& 4, characterized by convectively unfavorable conditions over the Bay of Bengal, also show weak OLR and wind anomalies over the Arabian Sea. This suggests that cyclonic conditions over the north Bay of Bengal might be vital in intensifying Arabian Sea systems --- as strong Arabian Sea anomalies are only observed in Regimes 1 \& 2 where cyclonic conditions persist over the Bay of Bengal.

\noindent A common feature of the first three regimes --- pronounced in Regimes 1 \& 2 and with a weaker short-lived signature in Regime 3 --- is the co-existence of a Bay of Bengal cyclonic anomaly during the development of the Arabian Sea system. In comparison, Regime 4 (and Regime 3, when mature) is characterized by an anticyclonic anomaly over the north Bay of Bengal. Interestingly, in all cases, the pattern in the Arabian Sea is poleward compared to the Bay of Bengal, which agrees with the northwest to the southeast orientation of the BSISO \citep{kikuchi2012bimodal}. Broadly, based on the evolution of these regimes and guided by previous work, the formation of systems over the Arabian Sea can be physically grouped into two classes, namely the downstream development of westward-moving lows that originated in the Bay of Bengal (Regime 1) and {\it in-situ} genesis (Regimes 2, 3 \& 4).

\section{Tracking of Western Indian Cyclonic Systems }

\noindent Though the lagged composite of clusters in Figure~\ref{fig:FIG4} provide some insight about weather patterns, identification of individual storms, their precursors, and evolution is now undertaken by tracking systems for 22 years (June-September; 1998-2019). Indeed, tracking of systems allows us to quantify local versus remotely formed systems based on their genesis, lysis, and trajectories. Further, this allows for an independent grouping of rainy systems in this region and thus also confirms the existence of clusters detected by the $k$-means approach. The tracking algorithm (described in the Method section) results in a total of 191 rainy \& long-lived cyclonic systems which pass over the Arabian Sea and western India (a marked red region in Figure S7). 
These systems are now classified based on the location of their genesis and the co-existence or absence of Bay of Bengal disturbances. Specifically, we define four categories as follows: (1) direct downstream intensification of westward-moving Bay of Bengal system (Type 1); any system falls in this category if its genesis is east of the region marked red in Figure S7 and lasts in the marked region at least 24 hours. This accounts for around 51\% or 98 systems; (2) {\it In-situ} formation over the Arabian Sea (Type 2); any system falls in this category if it forms over the Arabian Sea and remains in the marked box at least for 24 hours; this accounts for about 48\% or 92 systems. Further, Type 2 is grouped into three sub-categories based on the presence or absence of a monsoon low to the east. In particular, Type 2a comprises of {\it in-situ} Arabian Sea system formation with a preceding Bay of Bengal system. This accounts for 31\% of systems. Type 2b is {\it in-situ} formation but which precedes a Bay of Bengal system. This accounts for 9-10\% of systems. Finally, Type 2c is {\it in-situ} formation in the absence of a system in the Bay of Bengal, and this accounts for the remaining 7-8\% of systems. For clarity, a flow chart of system classification is shown in Figure S8. Thus, for 22 monsoon seasons, about 90\% of the synoptic systems observed in the Arabian Sea and western Indian region are associated with cyclonic anomalies (either preceding or after) in the Bay of Bengal. This is consistent with previous case studies in that stand-alone formation of rainy systems over the Arabian sea is relatively rare, and most of the system formation occurs with the Bay of Bengal low-pressure system. Furthermore, it is comforting to note that the classification of systems based on their genesis and tracks into Type 1, 2a, 2b \& 2c matches, at least qualitatively, with the patterns (Regimes 1, 2, 3 \& 4) obtained via unsupervised $k$-means clustering.

\subsection{Evolution of four categories}

\begin{figure*}
\centering
\includegraphics[trim=0 0 0 0, clip,height = 1\textwidth,width = 0.6\textwidth, angle =90, clip]{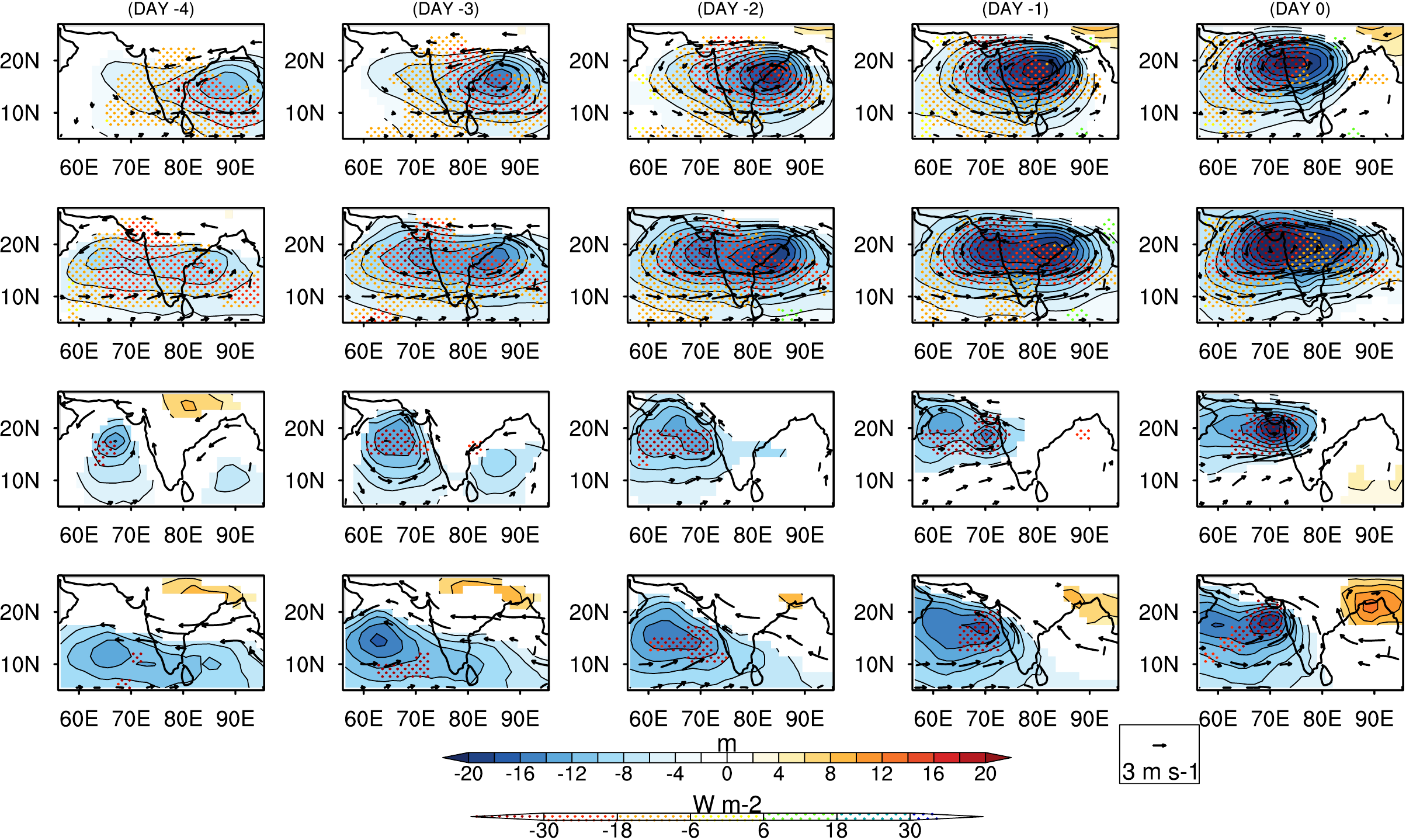}
\caption{Lag composites of four variants of Arabian Sea systems via cyclone tracking from Day $-4$ to Day 0. Rows 1, 2, 3 and 4 represent Type 1, 2a, 2b and 2c, respectively. Zero days represent the day of the heaviest rainfall in western India. Color shading represents the composite 600 hPa height anomaly (m), dotted shading is OLR anomaly, and arrows represent the composite wind anomaly. Height anomaly and OLR have shown only if they are significantly different from climatology at 0.1 level of significance and arrows if any wind component is significantly different from zero at 0.1 significance.} 
\label{fig:FIG5}
\end{figure*}

\noindent The lag-composites of 600 hPa height, OLR, and wind anomalies for the four types of MTCs identified by tracking of cyclonic systems over 22 years (Type1, Type 2a, 2b \& 2c) are shown in Figure~\ref{fig:FIG5}. In Type 1 composite evolution (first row; Figure~\ref{fig:FIG5}), initially, a height depression, cyclonic circulation, and associated negative OLR anomaly develop over the Bay of Bengal. This occurs at least four days before the onset and intensification of the system over the Arabian Sea. A westwards extending trough and weak negative OLR anomaly along the west coast of India are also observed prior to system formation. Gradually, the cyclonic circulation over the Bay intensifies and moves northwestwards, which finally, around Day $-1$ approaches western India. This category of system formation was found to be most frequent (98 systems), and consistent with previous work, suggests that, quite often, Arabian Sea systems form from the downstream evolution of westward-moving Bay of Bengal lows \citep{carr1977mid,ksn}. Given that this formation mechanism involves direct downstream development, its frequency is expected to be highly sensitive to the formation of cyclonic systems in the Bay of Bengal. The evolution of this type is qualitatively similar to Regime 1 of $k$-means clustering.

\noindent The composite evolution of Type 2a formation (second row; Figure~\ref{fig:FIG5}) shows that these systems develop locally over the Arabian Sea but in the presence of a monsoonal disturbance over the Bay of Bengal or East India. Four days before the intensification of the Arabian Sea system, a relatively strong cyclonic circulation and height anomaly is observed to develop over the Bay of Bengal, along with a zonally oriented trough and associated negative height anomaly over the Arabian Sea. A negative OLR anomaly accompanies this along the Western Ghats and over the Bay of Bengal in the southwest sector of cyclonic circulation. From Day $-4$ to Day $-3$, the Bay of Bengal height anomaly deepens, convection intensifies, and both systems encircle each other. This is immediately followed by an intensification of the Arabian Sea cyclonic system. Indeed, from Day $-3$ to Day 0, the Arabian Sea system remains stationary and intensifies significantly. Eventually, from Day $-1$ to Day $-0$, the Bay of Bengal system weakens and almost merges with the Arabian Sea system. Notably, the evolution is qualitatively similar to Regime 2 of $k$-means clustering. 

\noindent Type 2b formation (third row; Figure~\ref{fig:FIG5}) occurs with weak cyclonic activity over the Bay of Bengal or Eastern India. However, contrary to Type 2a, the genesis of this class of systems precedes the system to the east. From Day $-4$ to $-3$, convection and cyclonic circulation appear over the southwest Arabian Sea, gradually moving northwards with a slight eastwards component. From Day $-3$ to $-2$, cyclonic shear and a negative OLR anomaly appear in the Bay of Bengal and East India. Interestingly after the weak negative height and OLR anomaly around Day $-3$ to Day $-2$, the eastern Bay of Bengal shows signs of positive height anomalies and anticyclonic shear. 
Gradually, from Day $-2$ to Day 0, the cyclonic anomaly in the Arabian Sea and associated convection moves northwards parallel to the west coast of India and remains stationary near the coast of Gujarat. Indeed, for Type 2b, when mature, the sense of circulation is opposite in the Arabian Sea and the north Bay of Bengal. Given the evolution of wind and OLR anomalies, this formation type depicts the prototype monsoon onset vortex, which usually helps the progress of monsoon over western India and the Arabian Sea in the early monsoon months \citep{krishnamurti1981onset}. The development of this Type 2b category is similar to Regime 3 of $k$-means clustering. 

\noindent Finally, Type 2c (fourth row; Figure~\ref{fig:FIG5}) is a category where the Arabian Sea system appears in the absence of convection and cyclonic circulation over the north Bay of Bengal. 
However, this is a case of northwestward propagation of an equatorial convective envelope into the Arabian Sea. In particular, on Day $-4$, a cyclonic flow and negative OLR anomalies existed over the southern Bay of Bengal, southern Arabian sea, and near Sri Lanka. These anomalies then propagate northwestward into the Arabian Sea from Day $-4$ to Day $-2$. Once in the Arabian Sea, the movement is predominantly northward along the western coast of India, and the formation of a cyclonic system follows in the northeast Arabian Sea. Notably, while the large-scale anomalies propagate northwestward from the southern Bay of Bengal, the cyclonic system develops locally in the Arabian Sea. Throughout the evolution of this system, positive height anomalies and anticyclonic circulation is observed over the core monsoon zone and north Bay of Bengal, similar to the Regime 4 of $k$-means clustering. 

\subsection{Meteorological characteristics of the four categories}

\begin{figure*}
\centering
\includegraphics[trim=0 0 0 0, clip,height = 1\textwidth,width = 0.5\textwidth, angle =0, clip]{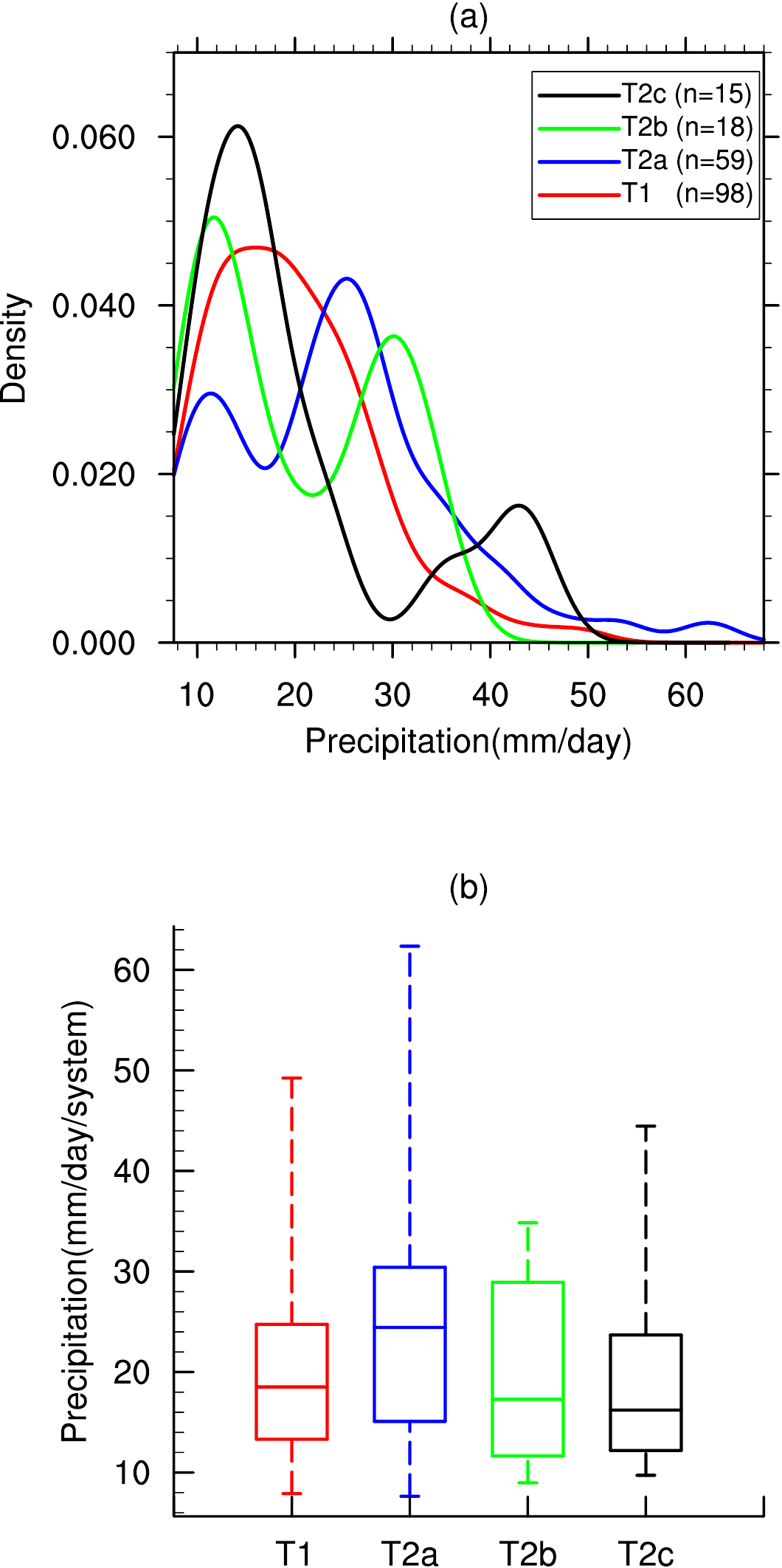}
\caption{Mean composite rainfall during cyclone life over the region ($18^{\circ}$N-$22^{\circ}$N $68^{\circ}$E-$72^{\circ}$E) (a) Gaussian Kernel Density Function (KDF) of mean composite rainfall; (b) box diagram of mean composite rainfall of four variants (Type 1, 2a, 2b and 2c).}
\label{fig:FIG6}
\end{figure*}

\begin{figure*}
\centering
\includegraphics[trim=0 0 0 0, clip,height = 1\textwidth,width = 1\textwidth, angle =0, clip]{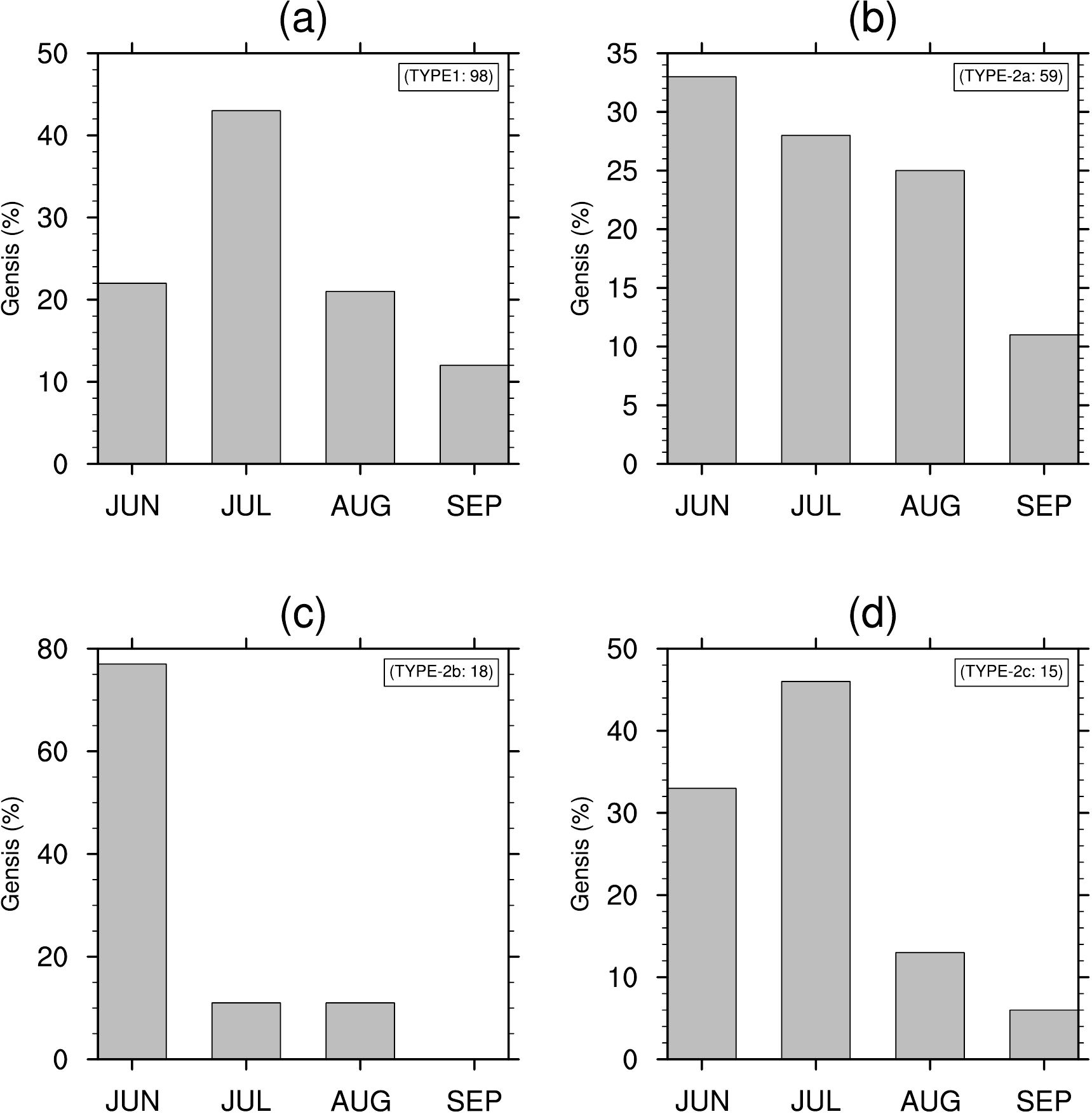}
\caption{Monthly frequency of occurrence of all four variants of Arabian Sea systems. (a) Type 1 ; (b) Type 2a; (c) Type 2b; (d) Type 2c.}
\label{fig:FIG7}
\end{figure*}

\begin{figure*}
\centering
\includegraphics[trim=0 0 0 0, clip,height = 0.8\textwidth,width = 1\textwidth, angle =0, clip]{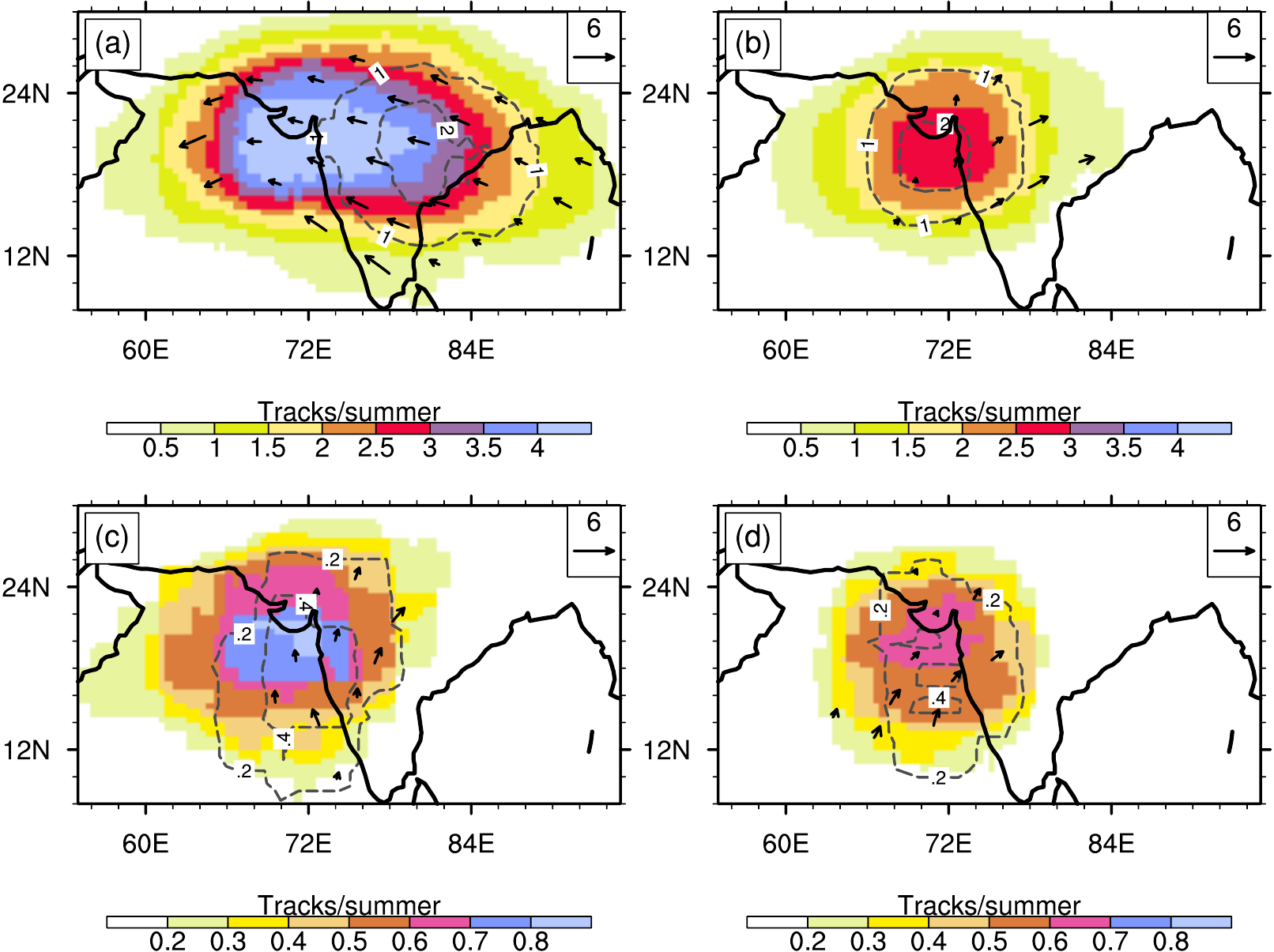}
\caption{Track density (color), mean cyclone motion vector (arrows) and cyclone genesis density (dashed contours) (a) Type 1, (b) Type 2a, (c) Type 2b, (d) Type 2c. Motion vector are shown if any of the propagation vector component is significantly different from zero at 0.1 significance.}
\label{fig:FIG8}
\end{figure*}

\noindent Having established that rainy MTCs over the Arabian Sea and western India fall into four categories with respect to the formation, the overall nature of which agrees with the $k$-mean classification scheme, we now examine various characteristics of these categories, including their rainfall distribution (Figure~\ref{fig:FIG6}), monthly frequency (Figure~\ref{fig:FIG7}), track density and motion vectors (Figure~\ref{fig:FIG8}). 

\noindent Type 1 formation, in terms of mean and extreme rainfall (Figure~\ref{fig:FIG6}), is the second-largest rain-producing category over western India and occurs most commonly in July and August (Figure~\ref{fig:FIG7}). The cyclone motion vectors and genesis density, shown in Figure~\ref{fig:FIG8}a, suggest that most of these systems take shape near the eastern state of Odisha in India, which is a little south of the head Bay of Bengal. As per the notion of their downstream development, motion vectors indicate that this category exhibits northwest movement across central India into the Arabian Sea. In contrast, Type 2a genesis occurs throughout the monsoon season (Figure~\ref{fig:FIG8}); however, this category is most frequent in June, and its occurrence monotonically decreases from June to September. The maximum track and genesis density of this class peaks in the northeast Arabian Sea suggests that these systems do not move much during their life cycle (Figure~\ref{fig:FIG8}b). Indeed, this is a major characteristic of the Arabian Sea synoptic systems during the monsoon \citep{carr1977mid,ksn}. 
As these systems remain close to the west coast of Maharashtra during their life cycle, their effect is expected to be more significant over the Indian landmass. Indeed, their slow motion and closeness to the west coast make them the rainiest (in terms of mean and extreme) type of systems experienced by western India (Figure~\ref{fig:FIG6}). In terms of rain rates, Type 2a systems reach intensities that exceed 60 mm/day, as is seen in the tail of the distribution in Figure~\ref{fig:FIG6}.

\noindent The Type 2b class moves northward parallel to the west coast of India with a slight eastward component towards the Indian landmass (Figure~\ref{fig:FIG8}c). This category's maximum track and genesis density are significantly less than Type 1 and Type 2a, and more spread out in a zonal direction. Notably, some systems of this class turn towards Indian landmass, thus having the potential of affecting western India, but overall their contribution to rainfall is lower than either Type 1 or Type 2a categories (Figure~\ref{fig:FIG6}). Further, in terms of the nature of heavy rainfall events, Type 2b is comparable to the Type 2a (Figure~\ref{fig:FIG6}a) and forms almost exclusively in June (Figure \ref{fig:FIG7}c), bearing a resemblance to the monsoon onset vortex \citep{krishnamurti1981onset,pearce1984onsets,yihui2005east}. Finally, Type 2c systems are the least frequent category forming mainly in July (Figure \ref{fig:FIG7}d). These systems progress northward along the west coast of India with motion towards the Indian landmass (Figure~\ref{fig:FIG7}d). As this class is weak and less frequent, it is overall less rain-bearing (Figure~\ref{fig:FIG6}). Interestingly, all the local formation types (Type2a, 2b \& 2c) show a bi-modal composite rainfall distribution, where the peaks are separated by almost $20-30$ mm/day. This suggests that though Types 2b \& 2c are less frequent and overall less rainy, systems in this category that correspond to the second peak may be hazardous. Further, it should be noted that individual rainfall events on a particular station would be might higher than what is presented in Figure~\ref{fig:FIG6} because it is a composite of precipitation over the life of the cyclone averaged over western India.


\section{Modulation by Intra-Seasonal Oscillations}

\begin{figure*}
\centering
\includegraphics[trim=0 0 0 0, clip,height = 1\textwidth,width = 0.5\textwidth, angle =0, clip]{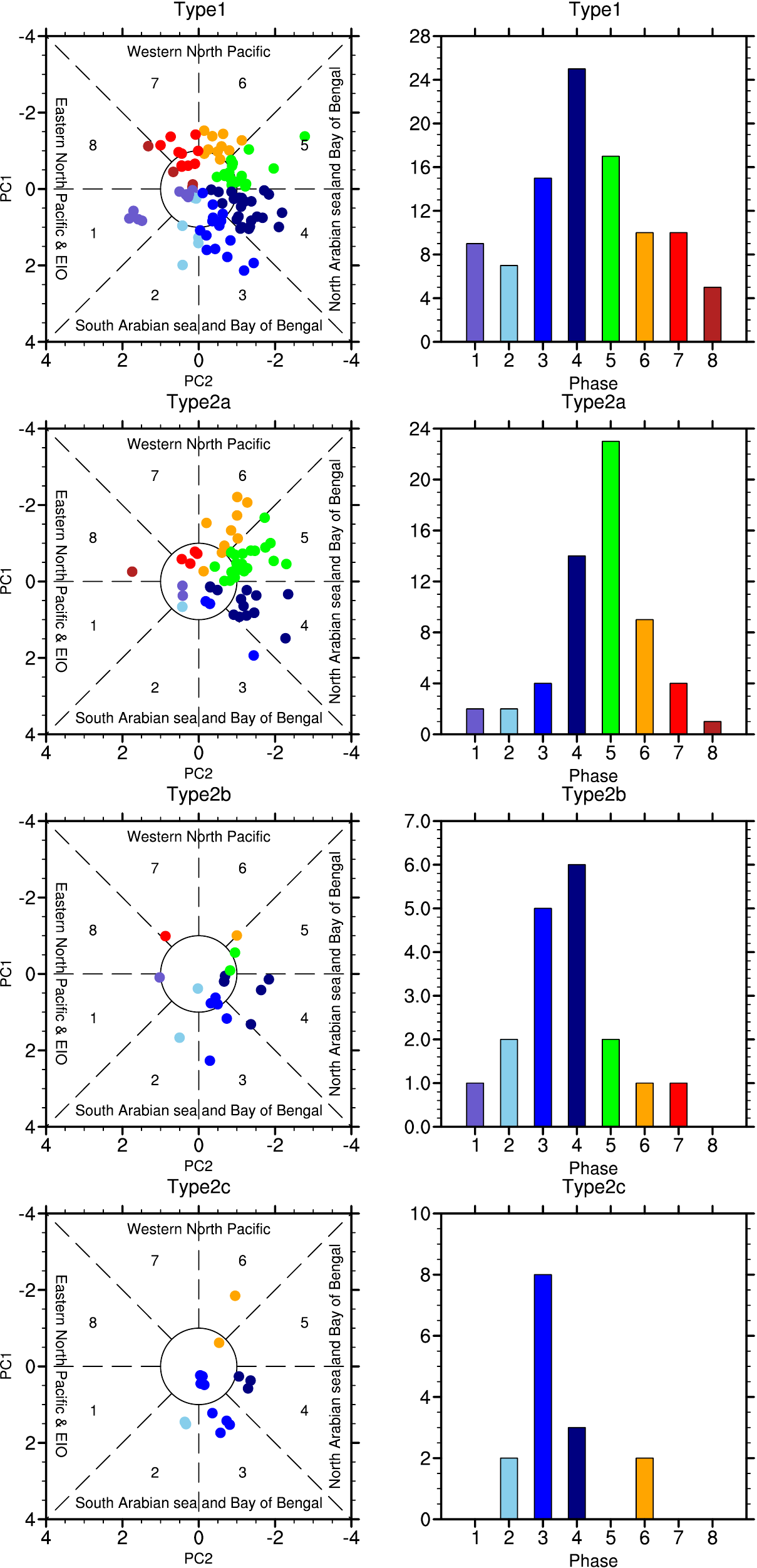}
\caption{Left Column: Phase diagram of BSISO during the genesis of each of the four variants of Arabian Sea systems; each dot represents the phase of BSISO during the genesis of a particular system of each category. Right column: Frequency distribution of phases of BSISO during the genesis of each category of synoptic systems.}
\label{fig:FIG9}
\end{figure*}

\noindent As seen in Figure \ref{fig:FIG1}, rainfall over western India and the northeast Arabian Sea does not happen in isolation; rather, it exists with moist convective activity over large regions similar to those seen in summer ISOs. Since ISOs can be effectively predicted on sub-seasonal time scales, this provides advance warning for systems that have a propensity for genesis in this large-scale environment. Hence, it is useful to explore if different categories of the Arabian Sea and western Indian systems identified so far have specific dependencies on indices that describe the activity of summer ISOs. Figure~\ref{fig:FIG9} shows the frequency of occurrence of each of the categories (Type 1, 2a, 2b \& 2c) with the phase and strength of BSISO during June-September, 1998-2019. In particular, the left column of Figure~\ref{fig:FIG9} represents the phase and magnitude of BSISO during the genesis of each system of the respective category, and the right column shows the corresponding frequency of formation during the respective phases of BSISO. Type 1 system formation occurs during all BSISO phases; however, the maximum frequency is during phases 4 and 5. This is not surprising because, during these phases, convection and cyclonic vorticity are enhanced over the north Bay of Bengal, which supports the formation of cyclonic systems \citep{karmakar2020differences,diaz2021evolution}. Since Type 1 systems over the Arabian Sea are a result of downstream development of Bay of Bengal lows, their frequency is expected to be directly modulated by cyclonic activity over East India and Bay of Bengal, thereby dependent on ISO phases which control low-pressure center formation over the Bay of Bengal \citep{deoras2021four,karmakar2021influence}.
Type 2a systems genesis also occurs during all phases of the ISO but shows peak activity during phase 5, followed by phase 4. Thus, Type 2a appears to share the same large-scale environment in terms of ISO phases as Type 1 for development. Moreover, in addition to the similarity with the phases 4 \& 5 of the BSISO, Type 2a composite during Days $-2$ to $-1$ (Figure~\ref{fig:FIG5}), exhibits remarkable similarity with the phases 2 \& 3 of the Quasi-Biweekly Oscillation \cite[QBWO; for example, Figure 10 in][]{qian2019new}.

\noindent Type 2b system formation peaks during phases 3 \& 4 of the BSISO, during which the maximum convection and cyclonic vorticity enhancement occur over the south-central Arabian Sea and west equatorial India Ocean. This is in accord with the recent work by \cite{deoras2021four,hunt2021modes} which suggests that LPS formation in the Arabian Sea and rainfall is usually enhanced during phases $2-4$ of the BSISO. Finally, Type 2c formation also peaks at phase $3$ of BSISO, when the convection over the equatorial Indian ocean is enhanced and migrates northward \citep{kikuchi2012bimodal}. An examination of the evolution of height anomaly and winds fields of Type 2c (Figure~\ref{fig:FIG5}) shows west-northwest motion tendency from the equator to the Arabian Sea during Day $-4$ to Day $-2$ and becomes predominately northwards from Day $-2$ to 0. Gradually, these height anomalies reach the northeast Arabian Sea and intensify into a cyclonic vortex. On Day $-4$, the height anomaly remains negative south of maximum easterly zonal winds (around 10-$15^{\circ}$N) and positive near central and north India (15-$25^{\circ}$N). This motion and height anomaly patterns closely resemble those of the QBWO \citep{chaterjee}. In fact, the QBWO is the dominant mode of variability at $15-25$ days time scales during the boreal summer and is known to modulate active and break phase of monsoon \citep{kikuchi2009global,qian2019new}. In fact, the influence of the QBWO on the formation of synoptic systems has been noted in various ocean basins \citep{ling2016impact,ghatak-sukhatme}. The positive height anomaly over central India and negative near the equator (Figure~\ref{fig:FIG5}) is consistent with phase 8 of QBWO \citep{chaterjee} and phases $1-3$ of BSISO \citep{kikuchi2012bimodal}, which correspond to a monsoon break over the north and central India, and an active spell over the southern part of the country. Thus, Type 2c represents a class of systems whose formation is triggered during the positive phase of BSISO and QBWO near the equator. 

\noindent Overall, these results indicate that instances of {\it in-situ} formation of Arabian Sea \& western Indian systems in the absence of Bay of Bengal cyclonic activity occurs when the ISOs (BSISO and/or QBWO) are in a favorable phase over this region. Moreover, the genesis of Type 1 and Type 2a, which is tied to cyclonic activity over the Bay of Bengal, is more efficient when the ISO is in an active phase over the north Bay of Bengal. Hence, the active phase of BSISO, and to some extent, the QBWO, and their location are essential for the Arabian Sea and western India system formation and evolution. This sensitivity of system formation on ISO phases is expected to some extent because positive phases of ISOs enhance large-scale moisture, vorticity, and barotropic instability, which are conditions that are required for the formation of synoptic monsoon systems \citep{karmakar2021influence}. 

\section{Conclusions}

\noindent Rain bearing synoptic systems over the northeast Arabian Sea and western India are analyzed by $k$-means clustering and cyclone tracking methodologies. Over twenty years of summer monsoon data from modern reanalysis is used to obtain a robust view of the systems responsible for a significant portion of the annual precipitation and extreme rainfall events in this part of the world. At the outset, a lag correlation analysis of OLR immediately shows a tendency for moist convection to simultaneously occur over the Arabian Sea and western Indian region and the Bay of Bengal. Further, the large-scale nature of the correlation patterns also suggests the likelihood that broader environmental conditions play a role in favoring the formation of rainy systems over this region.

\noindent Dominant patterns during rainy days over the Arabian Sea and western Indian are extracted by a $k$-means analysis which is performed using an optimum of four clusters, as decided by the silhouette coefficient and elbow methods. Each of the four clusters is obtained using daily 600 hPa geopotential height anomaly as input data vectors of western India's rainy days. Physically, all four clusters represent different weather regimes consisting of middle tropospheric cyclonic circulation anomalies over the northeast Arabian Sea and western India.
The height anomalies of the first regime have a northwest orientation and range from the Arabian Sea to the Bay of Bengal. Lagged composites indicate that synoptic systems over Arabian Sea in this regime develop from the westward movement of monsoon lows over the Bay of Bengal. Regime two shows concomitant cyclonic activity over the Arabian Sea and Bay of Bengal and suggests dynamical interaction between the two circulation patterns. In the third regime, systems form locally in the South Central Arabian Sea and move northwards with a weak transient cyclonic signature in the Bay of Bengal. 
Further, lagged composites indicate that the Bay of Bengal system intensification precedes or follows the Arabian Sea system genesis in the second and third regimes, respectively. In fact, in the third regime, when the Arabian Sea system matures, the circulation in the Bay is anticyclonic. The fourth regime consists of a large-scale cyclonic envelope that gradually moves northwest from the South Bay of Bengal and leads to the formation of a synoptic system in the Arabian Sea. 

\noindent With this broad characterization in four clusters, a detailed cyclone tracking methodology is used to track the systems and follow their development. In particular, 191 rainy synoptic systems that occur over the northeast Arabian Sea and western India in 22 years are tracked and classified. The resulting classification brings out four categories: the first (Type 1) accounts for 51\% of all systems, and these form from the westward (downstream) development of cyclonic lows over the Bay of Bengal. Second, Type 2a accounts for 31\% cases and consists of the formation of MTCs over the Arabian Sea with a coexisting Bay of Bengal system that precedes the Arabian Sea system formation. Third, Type 2b (9-10\%) formation is again characterized by the coexisting cyclonic circulations over the two basins. However, in this category, the Arabian Sea system precedes a relatively weak and short-lived cyclonic circulation over the Bay of Bengal. Finally, Type 2c accounts for the remaining 7-8\% of systems that form locally in the Arabian Sea from a large-scale cyclonic envelope that propagates northwestward from the southern Bay of Bengal. It is noteworthy that no {\it a priori} constraint was imposed on the tracking procedure, and it too yields four categories that qualitatively match the patterns captured by the $k$-means approach.

\noindent Tracking also allows for a study of motion vectors associated with synoptic systems, allows quantification of systems rainfall, and the monthly frequency of genesis of different categories of systems. We observe that {\it in-situ} genesis (specifically, Types 2a \& 2b) is favored in the early part of the monsoon season (i.e., June). In contrast, downstream development, Type 1 and Type 2c (i.e., {\it in-situ} Arabian Sea systems triggered from a large-scale cyclonic anomaly of south Bay of Bengal) are most frequently observed in July. Among all categories, Type 2a genesis occurs throughout the monsoon (June-September) and is rainiest, with the highest rain rates in western India that exceed 60 mm/day.
Further, the motion vectors of Type 2a suggest a marked quasi-stationary nature --- a hallmark of Arabian Sea MTCs --- when over the northeast Arabian Sea and western India, this leads to a large quantum of rain and potential flooding in the continental portion of this region. The westward-moving category (Type 1) moves off from western India to the Arabian Sea, but given its track, it accounts for the second rainiest system in this region. Interestingly, Type 2b, where the Arabian Sea system precedes the Bay of Bengal system, motion vectors indicate the possibility of curving into the Indian landmass, thus though much less frequent than Types 1 \& 2a, they too have an influence on rainfall in the western coastal regions of India. Finally, Type 2c systems usually progress northward into the eastern Arabian Sea and at times make landfall over the west coast; however, they contribute the least out of the four categories to rainfall in western India. 

\noindent With regard to the large-scale environment, it is seen that each of the four categories is preferentially formed in certain phases of the BSISO, which is the dominant mode of intraseasonal variability in the Indian region during the summer monsoon. In particular, Types 1 \& 2a, both of which require the prior presence of a cyclonic system over the Bay of Bengal, are most active in phases 4 and 5 of the BSISO. Indeed, these are precisely the phases of the BSISO when moist convection is favored over the Bay of Bengal. Phases 3 and 4 of BSISO, when convection is active in the western equatorial Indian Ocean, and the central Arabian Sea, are favorable for the genesis of the categories where the Arabian Sea system precedes cyclonic circulation over the Bay (Type 2b) and when the Arabian Sea system develops from the northwest movement of large-scale cyclonic anomalies from the South Bay of Bengal (Type 2c). Apart from an association with the BSISO, Type 2a \& 2c systems also show a preference for particular phases of the QBWO; in particular, the large-scale envelope from which Type 2c systems are born closely resembles the boreal summer QBWO. Thus, both intraseasonal modes, the BSISO and the QBWO appear to aid the formation of synoptic systems over the Arabian Sea and western India.

\noindent In all, from the lag-correlation patterns of OLR, the regimes from $k$-means clustering and tracking systems over twenty monsoon seasons, it is clear that most rainy synoptic systems in the Arabian Sea and western India form with, or from, cyclonic anomalies in the north Bay of Bengal. Indeed, more than 90\% of rain-bearing systems in this region fall in Type 1 and Types 2a \& 2b categories. Even Type 2c, where the north Bay of Bengal shows unfavorable conditions for the moist cyclonic activity, forms from a large-scale cyclonic anomaly that propagates into the Arabian Sea from the southern Bay of Bengal. The qualitative consistency in the clusters found by $k$-means and the groups delineated by cyclone tracking is an encouraging indicator of the robustness of the four basic regimes that encompass much of the rainy spells experienced in western India. We are hopeful that these regimes, their precursors, and the identification of large-scale intraseasonal phases in which they occur will help in the potential predictability and preparedness for extreme events associated with these systems. Finally, we note that the dynamical interaction between cyclonic activity over the Bay of Bengal and the formation of Arabian Sea MTCs will be presented in a companion paper.

\clearpage

\begin{center}
{\Large {\bf Supplementary Material}}
\end{center}

\begin{figure}
\includegraphics[trim=0 0 0 0, clip,height = 0.8\textwidth,width = 1\textwidth, angle =0, clip]{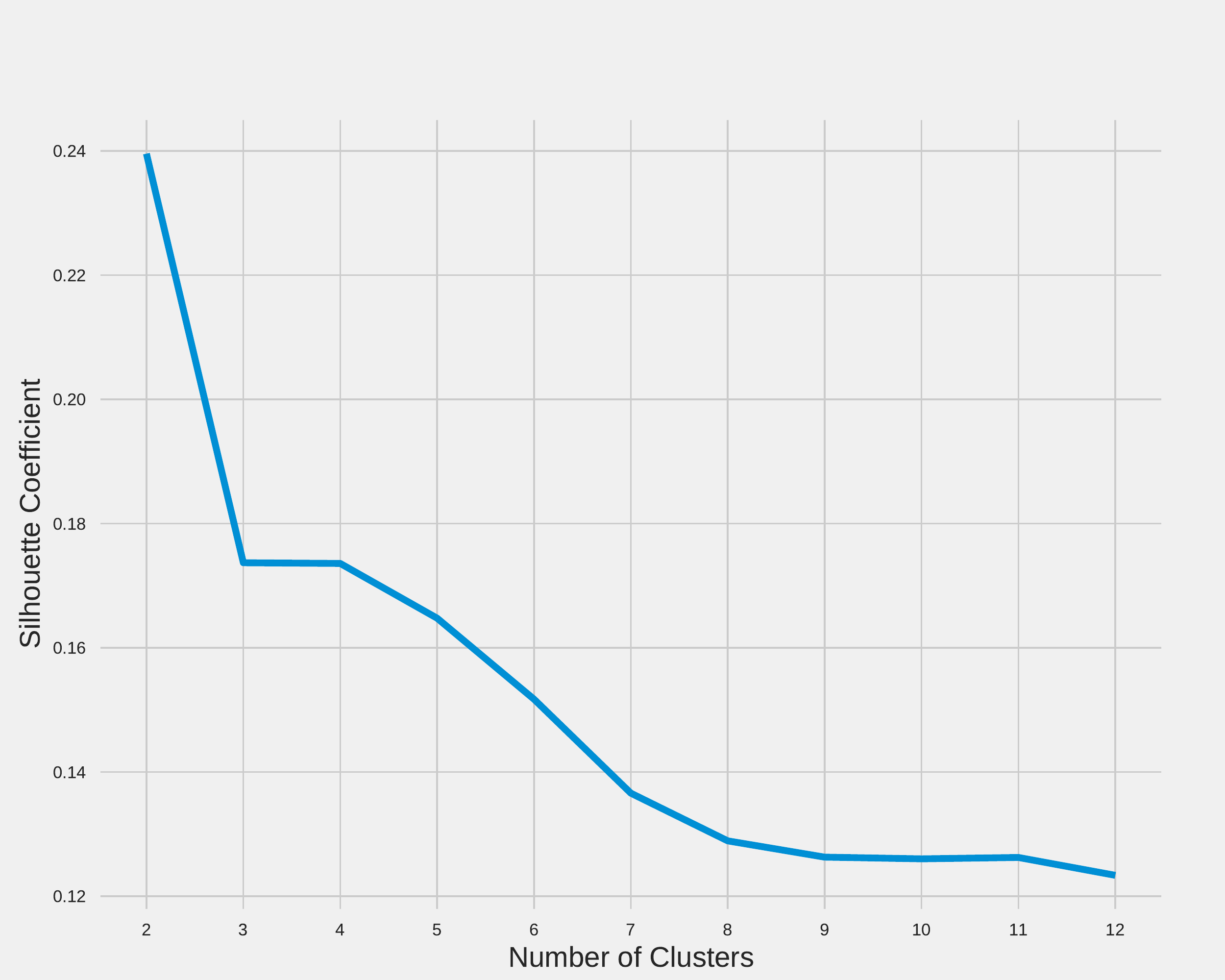}
\caption{Figure show average Silhouette Coefficient vs number of clusters (k)}
\label{FIG:S1}
\end{figure}

\begin{figure}
\includegraphics[trim=0 0 0 0, clip,height = 0.8\textwidth,width = 1\textwidth, angle =0, clip]{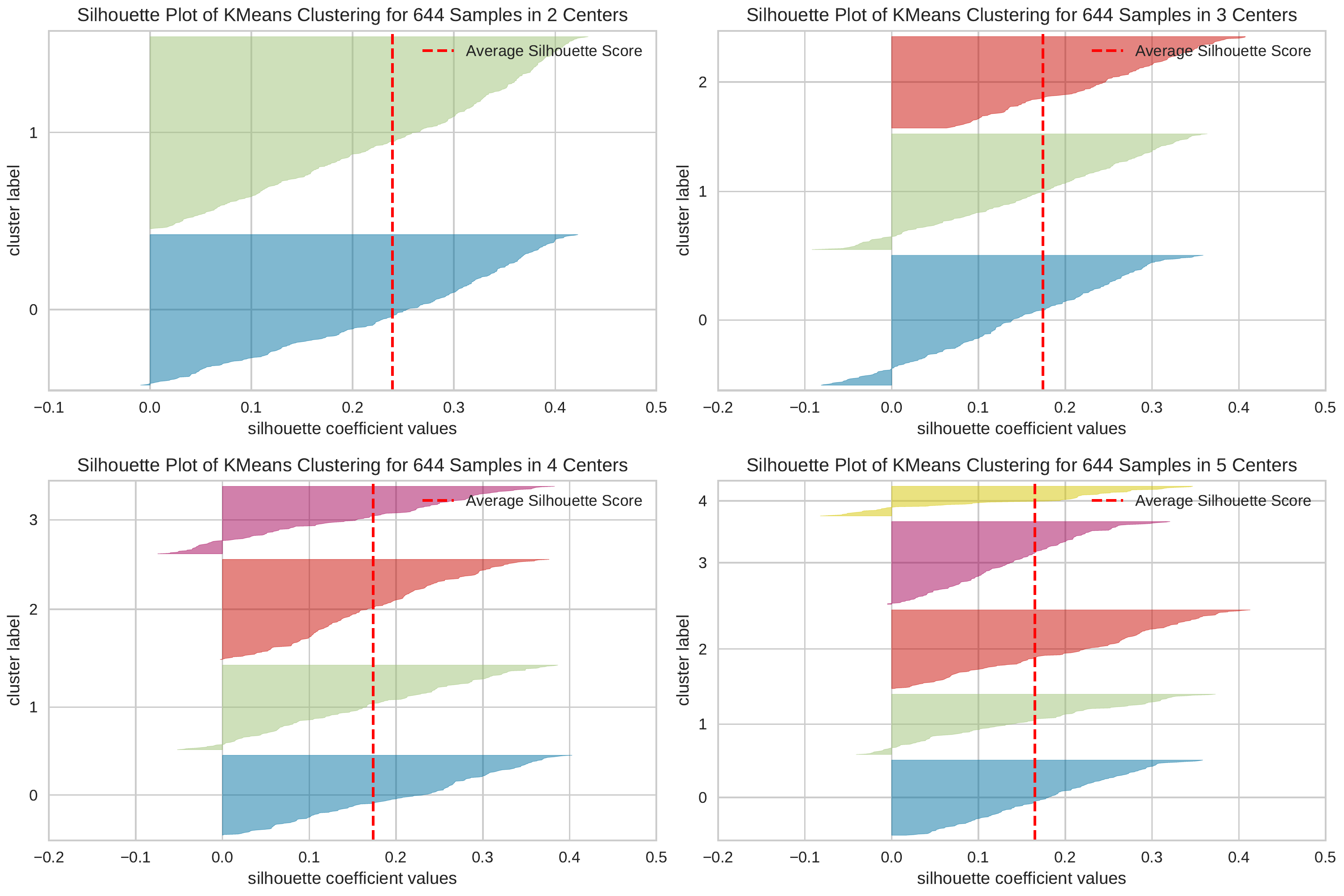}
\caption{Figure show Silhouette Coefficient for respected clusters vs number of clusters (k).}
\label{FIG:S2}
\end{figure}

\begin{figure}
\includegraphics[trim=0 0 0 0, clip,height = 0.95\textwidth,width = 1\textwidth, angle =0, clip]{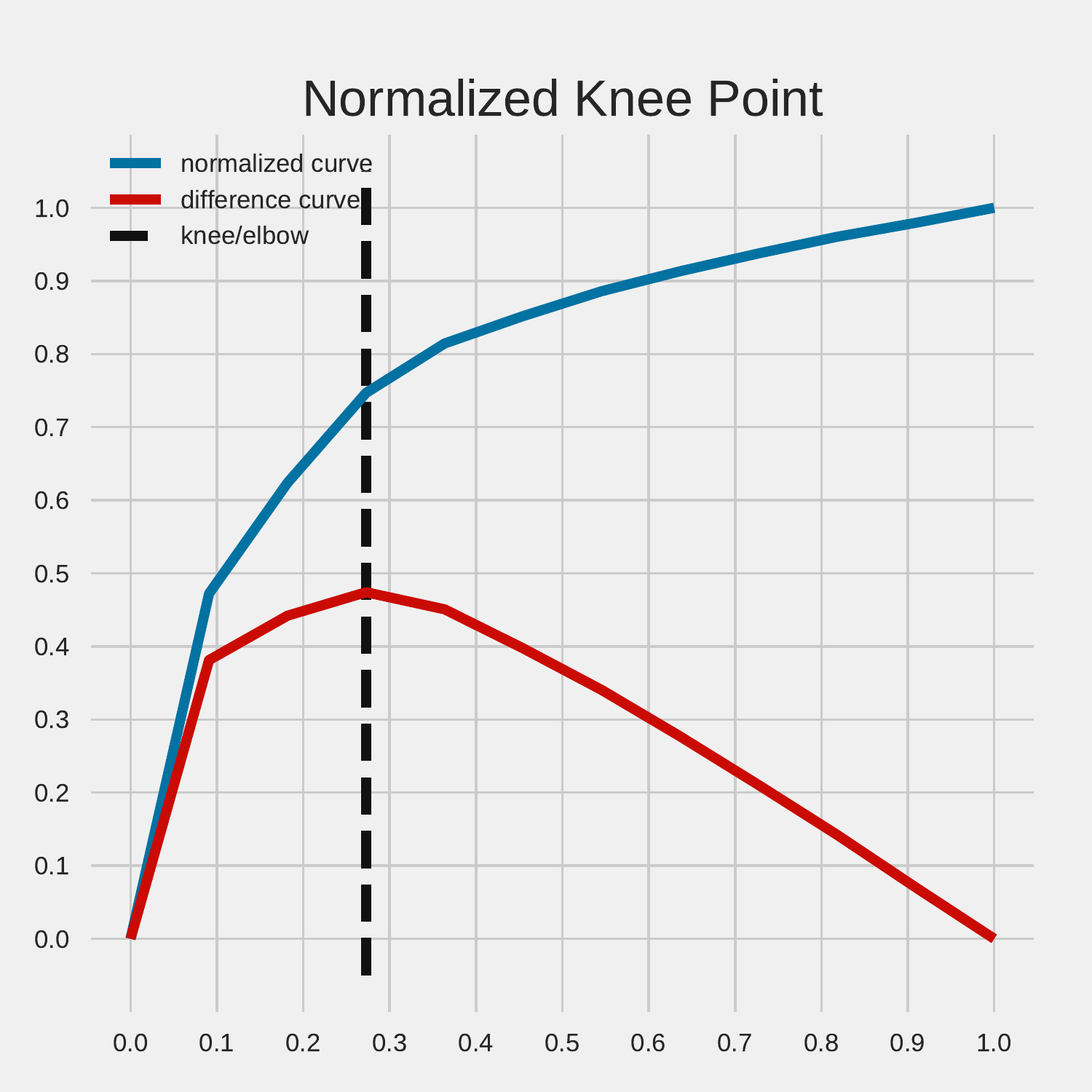}
\caption{Figure show the normalized error, difference and knee/elbow point; based on Satopaa et al.,2011}
\label{FIG:S3}
\end{figure}

\begin{figure}
\includegraphics[trim=0 0 0 0, clip,height = 1\textwidth,width = 1.0\textwidth, angle =0, clip]{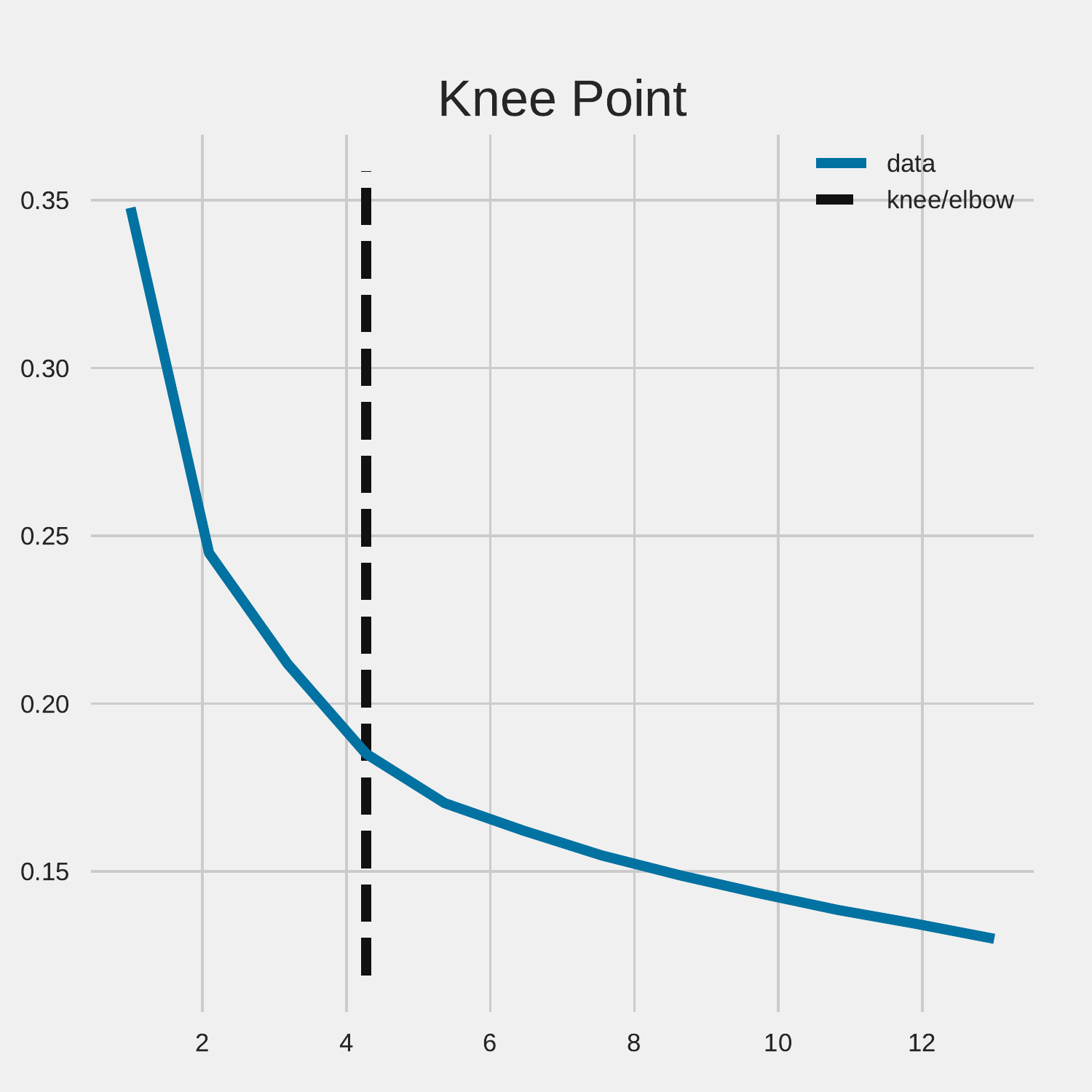}
\caption{Total within cluster sum of square distances (WSS$\times 10^{-5}$) vs number of clusters k; black line represent the knee point (k = 4.27) calculated using kneedle (Satopaa et al., 2011),. }
\label{FIG:S4}
\end{figure}

\begin{figure}
\includegraphics[trim=0 0 0 0, clip,height = 0.95\textwidth,width = 0.8\textwidth, angle =0, clip]{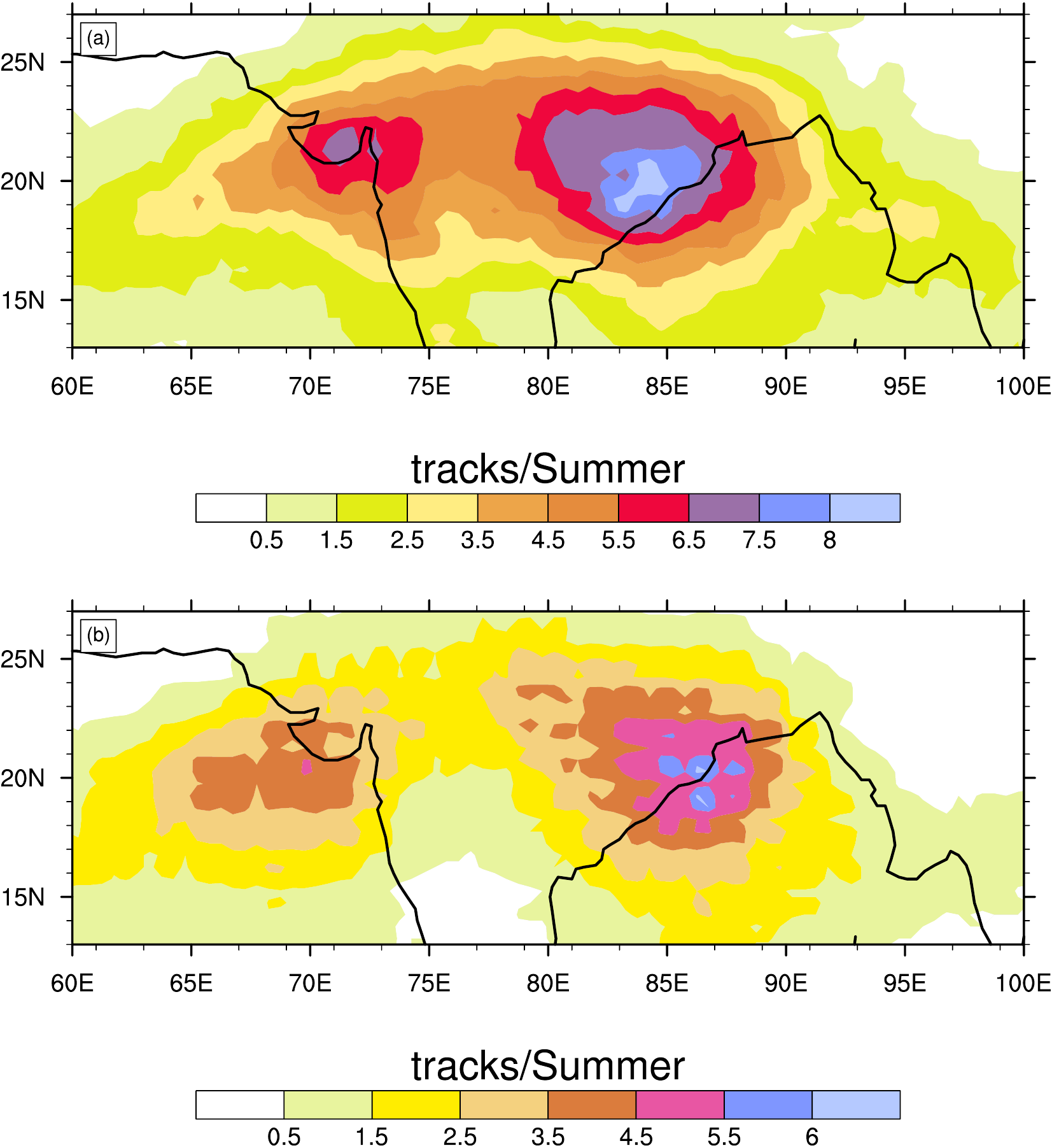}
\caption{(a) Track density of automatic method (b) and of manually extended tracks of \citep{ksn}; Variables are density over $3^{\circ}\times3^{\circ}$ grid.}
\label{FIG:S5}
\end{figure}

\begin{figure}
\includegraphics[trim=0 0 0 0, clip,height = 1.3\textwidth,width = 1\textwidth, angle =0, clip]{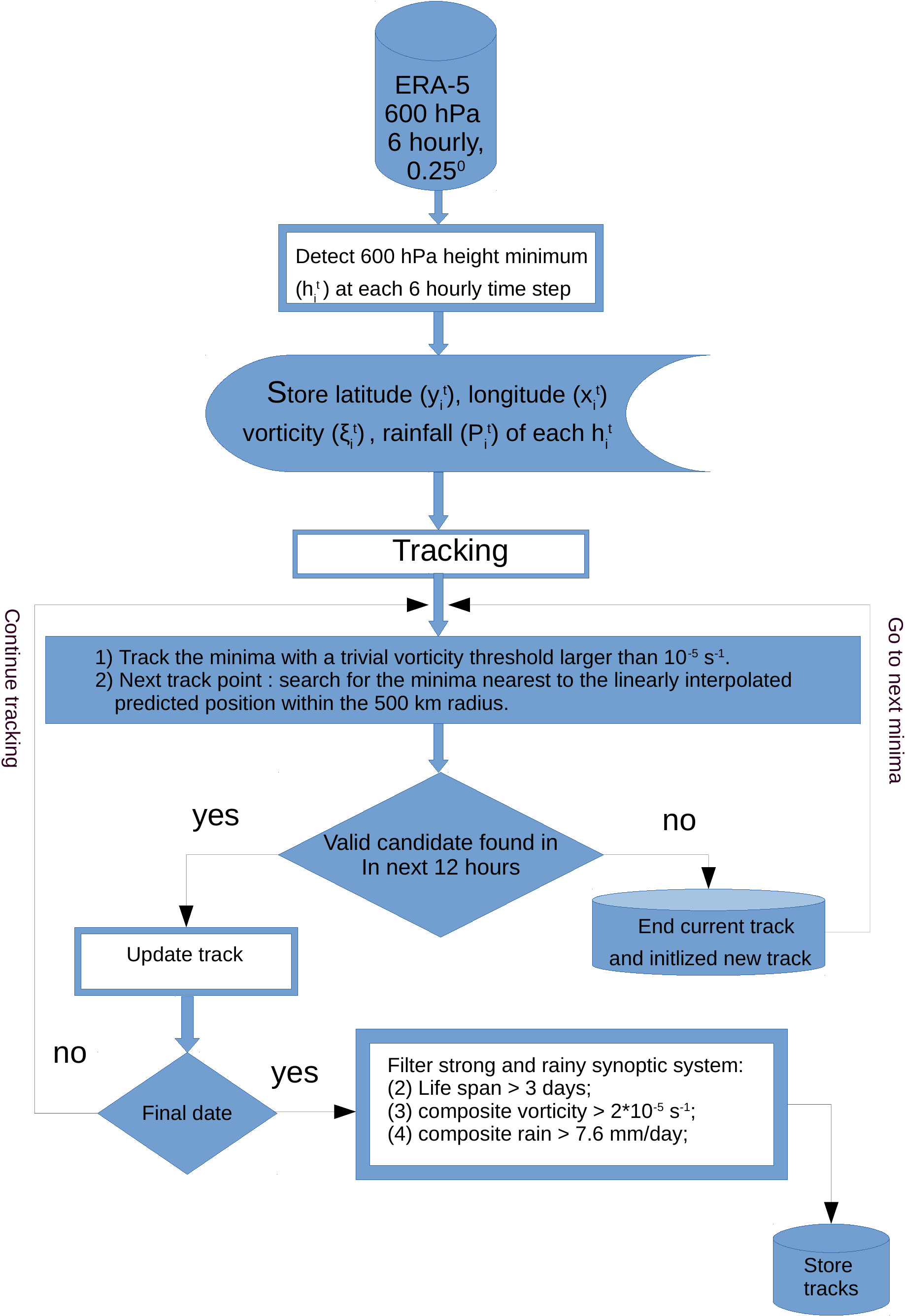}
\caption{Schematic of tracking procedure}
\label{fig:S6}
\end{figure}

\begin{figure}
\includegraphics[trim=0 0 0 0, clip,height = 0.6\textwidth,width = 0.9\textwidth, angle =0, clip]{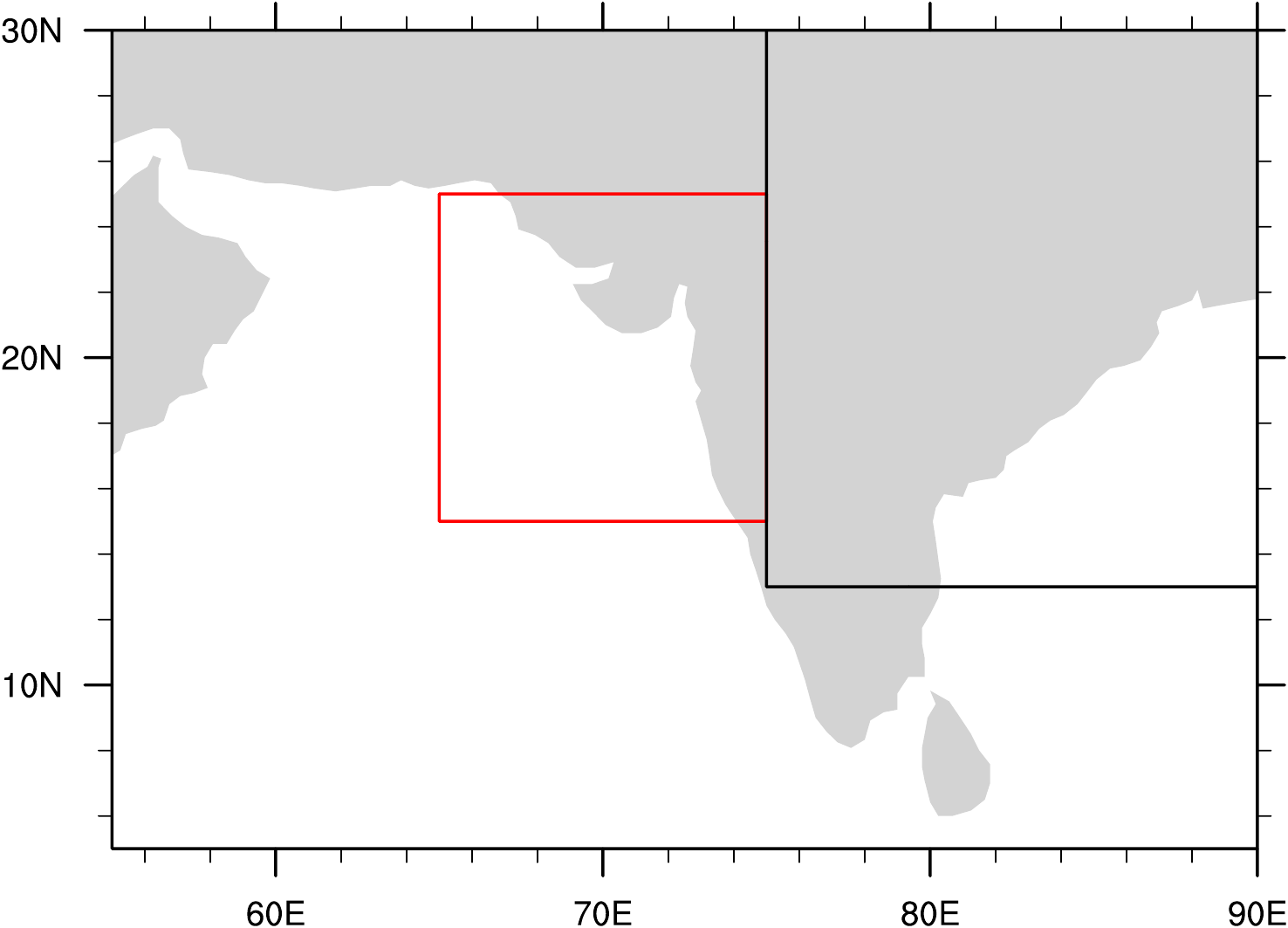}
\caption{Domains Specifications: Red box: Arabian sea; Black box: Bay of Bengal}
\label{fig:S7}
\end{figure}

\begin{figure}
\includegraphics[trim=0 0 0 0, clip,height = 1.3\textwidth,width = 1\textwidth, angle =0, clip]{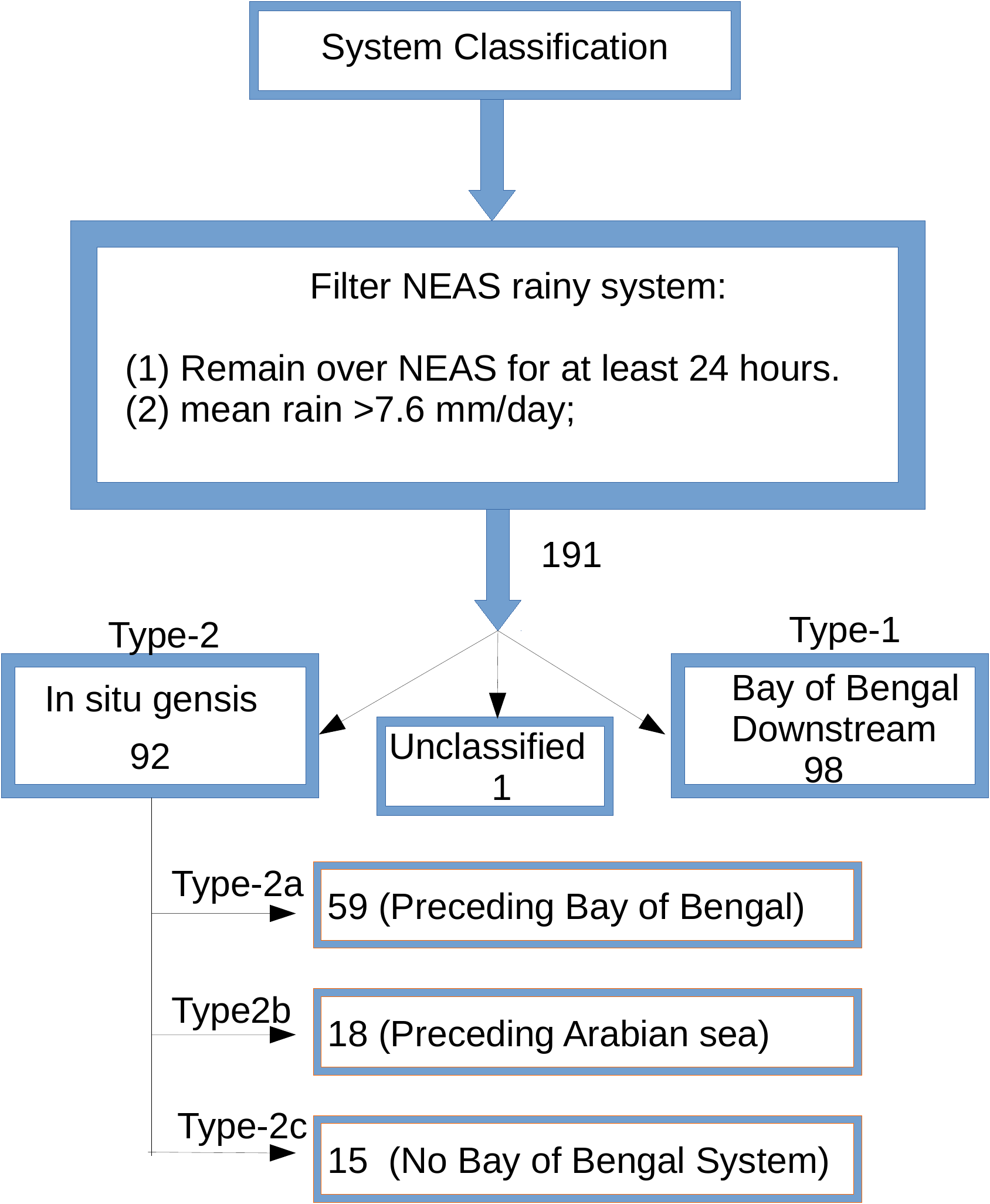}
\caption{Arabian Sea system classification}
\label{fig:S8}
\end{figure}

\begin{figure}
\includegraphics[trim=0 0 0 0, clip,width = 1\textwidth,height = 0.8\textwidth, angle =0, clip]{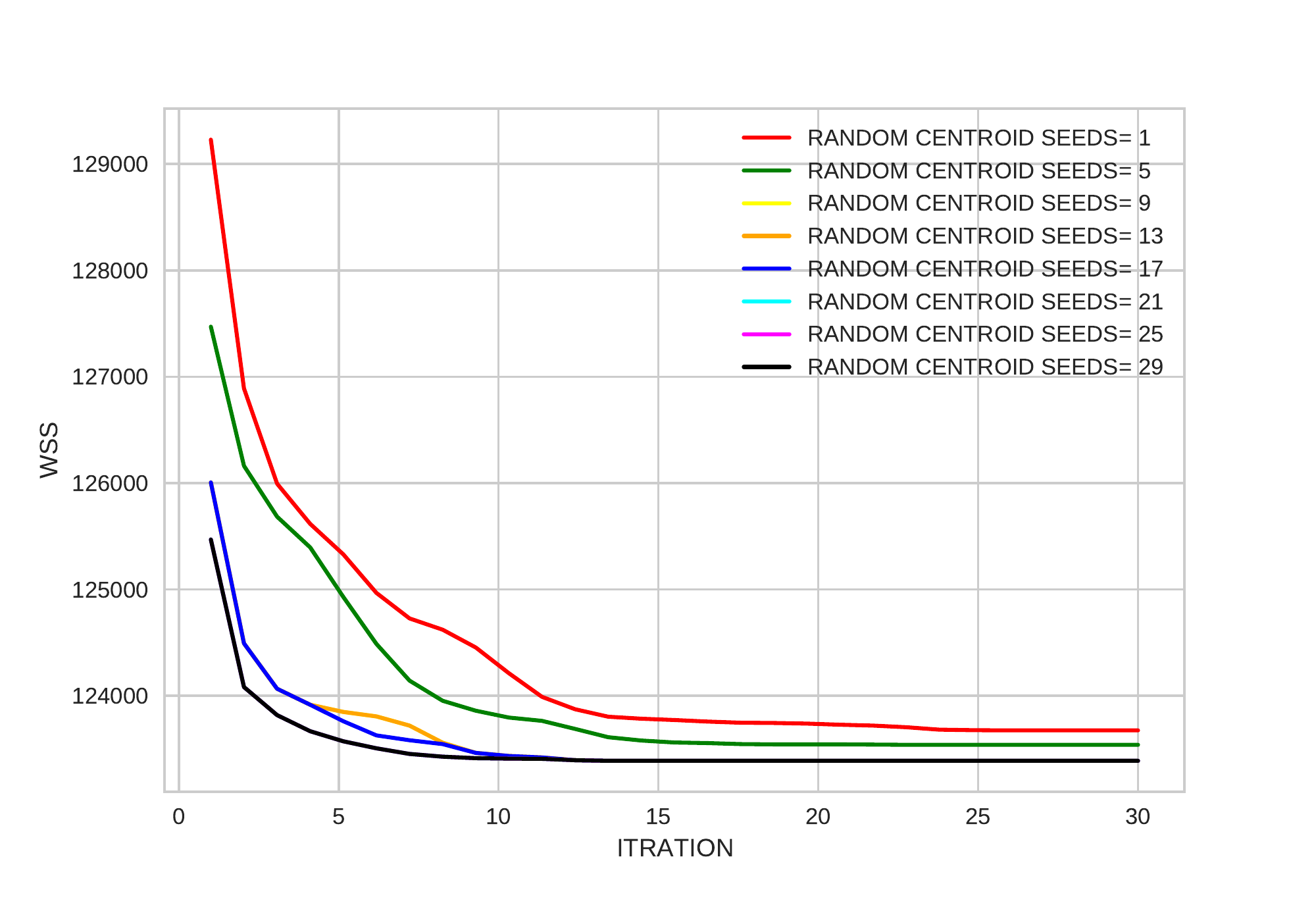}
\caption{Convergence of K-means; Variation of total within cluster sum of squares with respect to the iteration. Different color represents initial random centroids used for initialization.}
\label{fig:S9}
\end{figure}

\newpage
\bibliographystyle{apalike}
\bibliography{ref.bib}

\begin{thebibliography}{}

\bibitem[Adames and Ming, 2018]{AdamesMing}
Adames, {\'A}. and Ming, Y. (2018).
\newblock Interactions between water vapor and potential vorticity in
  synoptic-scale monsoonal disturbances: Moisture vortex instability.
\newblock {\em Journal of the Atmospheric Sciences}, 75(6):2083--2106.

\bibitem[Awan et~al., 2015]{awan2015identification}
Awan, J.~A., Bae, D.-H., and Kim, K.-J. (2015).
\newblock Identification and trend analysis of homogeneous rainfall zones over
  the east asia monsoon region.
\newblock {\em International Journal of Climatology}, 35(7):1422--1433.

\bibitem[Bholowalia and Kumar, 2014]{bholowalia2014ebk}
Bholowalia, P. and Kumar, A. (2014).
\newblock Ebk-means: A clustering technique based on elbow method and k-means
  in wsn.
\newblock {\em International Journal of Computer Applications}, 105(9).

\bibitem[Bian et~al., 2021]{bian2021well}
Bian, G.-F., Nie, G.-Z., and Qiu, X. (2021).
\newblock How well is outer tropical cyclone size represented in the era5
  reanalysis dataset?
\newblock {\em Atmospheric Research}, 249:105339.

\bibitem[Boos et~al., 2015]{boos2015adiabatic}
Boos, W., Hurley, J., and Murthy, V. (2015).
\newblock Adiabatic westward drift of indian monsoon depressions.
\newblock {\em Quarterly Journal of the Royal Meteorological Society},
  141(689):1035--1048.

\bibitem[Brode and Mak, 1978]{brode1978mechanism}
Brode, R. and Mak, M. (1978).
\newblock On the mechanism of the monsoonal mid-tropospheric cyclone formation.
\newblock {\em Journal of the Atmospheric Sciences}, 35(8):1473--1484.

\bibitem[Carr, 1977]{carr1977mid}
Carr, F.~H. (1977).
\newblock Mid-tropospheric cyclones of the summer monsoon.
\newblock {\em pure and applied geophysics}, 115(5-6):1383--1412.

\bibitem[Chaterjee and Goswami, 2004]{chaterjee}
Chaterjee, P. and Goswami, B. (2004).
\newblock {Structure, genesis and scale selection of the tropical
  quasi-biweekly mode}.
\newblock {\em QJRMS}.

\bibitem[Chen and Weng, 1999]{chen1999interannual}
Chen, T.-C. and Weng, S.-P. (1999).
\newblock Interannual and intraseasonal variations in monsoon depressions and
  their westward-propagating predecessors.
\newblock {\em Monthly Weather Review}, 127(6):1005--1020.

\bibitem[Choudhury et~al., 2018]{choudhury2018phenomenological}
Choudhury, A.~D., Krishnan, R., Ramarao, M., Vellore, R., Singh, M., and Mapes,
  B. (2018).
\newblock A phenomenological paradigm for midtropospheric cyclogenesis in the
  indian summer monsoon.
\newblock {\em Journal of the Atmospheric Sciences}, 75(9):2931--2954.

\bibitem[Clark et~al., 2018]{clark2018rainfall}
Clark, S., Reeder, M.~J., and Jakob, C. (2018).
\newblock Rainfall regimes over northwestern australia.
\newblock {\em Quarterly Journal of the Royal Meteorological Society},
  144(711):458--467.

\bibitem[Deoras et~al., 2021]{deoras2021four}
Deoras, A., Hunt, K., and Turner, A. (2021).
\newblock The four varieties of south asian monsoon low-pressure systems and
  their modulation by tropical intraseasonal variability.
\newblock {\em Weather}, 76(6):194--200.

\bibitem[Diaz and Boos, 2019a]{diaz2019barotropic}
Diaz, M. and Boos, W.~R. (2019a).
\newblock Barotropic growth of monsoon depressions.
\newblock {\em Quarterly Journal of the Royal Meteorological Society},
  145(719):824--844.

\bibitem[Diaz and Boos, 2019b]{diaz2019monsoon}
Diaz, M. and Boos, W.~R. (2019b).
\newblock Monsoon depression amplification by moist barotropic instability in a
  vertically sheared environment.
\newblock {\em Quarterly Journal of the Royal Meteorological Society},
  145(723):2666--2684.

\bibitem[Diaz and Boos, 2021]{diaz2021evolution}
Diaz, M. and Boos, W.~R. (2021).
\newblock Evolution of idealized vortices in monsoon-like shears: Application
  to monsoon depressions.
\newblock {\em Journal of the Atmospheric Sciences}, 78(4):1207--1225.

\bibitem[Dwivedi et~al., 2021]{dwivedi2021variability}
Dwivedi, S., Yesubabu, V., Ratnam, M.~V., Dasari, H.~P., Langodan, S., Raj,
  S.~A., and Hoteit, I. (2021).
\newblock Variability of monsoon inversion over the arabian sea and its impact
  on rainfall.
\newblock {\em International Journal of Climatology}, 41:E2979--E2999.

\bibitem[Francis and Gadgil, 2006]{francis2006intense}
Francis, P. and Gadgil, S. (2006).
\newblock Intense rainfall events over the west coast of india.
\newblock {\em Meteorology and Atmospheric Physics}, 94(1-4):27--42.

\bibitem[Ghatak and Sukhatme, 2022]{ghatak-sukhatme}
Ghatak, S. and Sukhatme, J. (2022).
\newblock Southwestward propagating quasi-biweekly oscillations over the
  south-west indian ocean during boreal winter.
\newblock {\em Weather and Climate Dynamics}.

\bibitem[Godbole, 1977]{godbole1977}
Godbole, R. (1977).
\newblock The composite structure of the monsoon depression.
\newblock {\em Tellus}, 29:25--40.

\bibitem[Goswami et~al., 1980]{goswami1980role}
Goswami, B., Keshavamurty, R., and Satyan, V. (1980).
\newblock Role of barotropic, baroclinic and combined barotropic-baroclinic
  instability for the growth of monsoon depressions and mid-tropospheric
  cyclones.
\newblock {\em Proceedings of the Indian Academy of Sciences-Earth and
  Planetary Sciences}, 89(1):79--97.

\bibitem[Goswami, 1987]{goswami1987mechanism}
Goswami, B.~N. (1987).
\newblock A mechanism for the west-north-west movement of monsoon depressions.
\newblock {\em Nature}, 326(6111):376--378.

\bibitem[Hanley and Caballero, 2012]{hanley2012objective}
Hanley, J. and Caballero, R. (2012).
\newblock Objective identification and tracking of multicentre cyclones in the
  era-interim reanalysis dataset.
\newblock {\em Quarterly Journal of the Royal Meteorological Society},
  138(664):612--625.

\bibitem[Hartigan and Wong, 1979]{hartigan1979ak}
Hartigan, J.~A. and Wong, M.~A. (1979).
\newblock Ak-means clustering algorithm.
\newblock {\em Journal of the Royal Statistical Society: Series C (Applied
  Statistics)}, 28(1):100--108.

\bibitem[Hersbach et~al., 2020]{hersbach2020era5}
Hersbach, H., Bell, B., Berrisford, P., Hirahara, S., Hor{\'a}nyi, A.,
  Mu{\~n}oz-Sabater, J., Nicolas, J., Peubey, C., Radu, R., Schepers, D.,
  et~al. (2020).
\newblock The era5 global reanalysis.
\newblock {\em Quarterly Journal of the Royal Meteorological Society},
  146(730):1999--2049.

\bibitem[Hodges et~al., 2003]{hodges2003comparison}
Hodges, K.~I., Hoskins, B.~J., Boyle, J., and Thorncroft, C. (2003).
\newblock A comparison of recent reanalysis datasets using objective feature
  tracking: Storm tracks and tropical easterly waves.
\newblock {\em Monthly Weather Review}, 131(9):2012--2037.

\bibitem[Huffman et~al., 2010]{huffman2010trmm}
Huffman, G.~J., Adler, R.~F., Bolvin, D.~T., and Nelkin, E.~J. (2010).
\newblock The trmm multi-satellite precipitation analysis (tmpa).
\newblock In {\em Satellite rainfall applications for surface hydrology}, pages
  3--22. Springer.

\bibitem[Huffman et~al., 2007]{huffman2007trmm}
Huffman, G.~J., Bolvin, D.~T., Nelkin, E.~J., Wolff, D.~B., Adler, R.~F., Gu,
  G., Hong, Y., Bowman, K.~P., and Stocker, E.~F. (2007).
\newblock The trmm multisatellite precipitation analysis (tmpa): Quasi-global,
  multiyear, combined-sensor precipitation estimates at fine scales.
\newblock {\em Journal of hydrometeorology}, 8(1):38--55.

\bibitem[Hunt et~al., 2016a]{Hunt}
Hunt, K., Turner, A., Innes, P., Parker, D., and Levine, R. (2016a).
\newblock {On the Structure and Dynamics of Indian Monsoon Depressions}.
\newblock {\em Monthly Weather Review}, 144:3391--3416.

\bibitem[Hunt and Fletcher, 2019]{hunt2019relationship}
Hunt, K.~M. and Fletcher, J.~K. (2019).
\newblock The relationship between indian monsoon rainfall and low-pressure
  systems.
\newblock {\em Climate Dynamics}, 53(3):1859--1871.

\bibitem[Hunt et~al., 2016b]{hunt2016spatiotemporal}
Hunt, K.~M., Turner, A.~G., and Parker, D.~E. (2016b).
\newblock The spatiotemporal structure of precipitation in indian monsoon
  depressions.
\newblock {\em Quarterly Journal of the Royal Meteorological Society},
  142(701):3195--3210.

\bibitem[Hunt et~al., 2021]{hunt2021modes}
Hunt, K.~M., Turner, A.~G., Stein, T.~H., Fletcher, J.~K., and Schiemann, R.~K.
  (2021).
\newblock Modes of coastal precipitation over southwest india and their
  relationship with intraseasonal variability.
\newblock {\em Quarterly Journal of the Royal Meteorological Society},
  147(734):181--201.

\bibitem[Hurley and Boos, 2015]{HurleyBoos}
Hurley, J. and Boos, W. (2015).
\newblock A global climatology of monsoon low-pressure systems.
\newblock {\em Quarterly Journal of the Royal Meteorological Society},
  141:1049--1064.

\bibitem[Janiga and Thorncroft, 2013]{janiga}
Janiga, M. and Thorncroft, C. (2013).
\newblock {Regional differences in the kinematic and thermodynamic structure of
  African easterly waves}.
\newblock {\em Quartely Journal of the Royal Meteorological Society},
  139:1598--1614.

\bibitem[Jiang et~al., 2016]{jiang2016global}
Jiang, N., Qian, W., and Leung, J. C.-H. (2016).
\newblock The global monsoon division combining the k-means clustering method
  and low-level cross-equatorial flow.
\newblock {\em Climate dynamics}, 47(7):2345--2359.

\bibitem[Karmakar et~al., 2021]{karmakar2021influence}
Karmakar, N., Boos, W.~R., and Misra, V. (2021).
\newblock Influence of intraseasonal variability on the development of monsoon
  depressions.
\newblock {\em Geophysical Research Letters}, 48(2):e2020GL090425.

\bibitem[Karmakar and Misra, 2020]{karmakar2020differences}
Karmakar, N. and Misra, V. (2020).
\newblock Differences in northward propagation of convection over the arabian
  sea and bay of bengal during boreal summer.
\newblock {\em Journal of Geophysical Research: Atmospheres},
  125(3):e2019JD031648.

\bibitem[Kikuchi, 2020]{kikuchi2020extension}
Kikuchi, K. (2020).
\newblock Extension of the bimodal intraseasonal oscillation index using jra-55
  reanalysis.
\newblock {\em Climate Dynamics}, 54(1):919--933.

\bibitem[Kikuchi and Wang, 2009]{kikuchi2009global}
Kikuchi, K. and Wang, B. (2009).
\newblock Global perspective of the quasi-biweekly oscillation.
\newblock {\em Journal of Climate}, 22(6):1340--1359.

\bibitem[Kikuchi and Wang, 2010]{kikuchi2010formation}
Kikuchi, K. and Wang, B. (2010).
\newblock Formation of tropical cyclones in the northern indian ocean
  associated with two types of tropical intraseasonal oscillation modes.
\newblock {\em Journal of the Meteorological Society of Japan. Ser. II},
  88(3):475--496.

\bibitem[Kikuchi et~al., 2012]{kikuchi2012bimodal}
Kikuchi, K., Wang, B., and Kajikawa, Y. (2012).
\newblock Bimodal representation of the tropical intraseasonal oscillation.
\newblock {\em Climate Dynamics}, 38(9):1989--2000.

\bibitem[Krishnamurthy and Ajayamohan, 2010]{krishnamurthy2010composite}
Krishnamurthy, V. and Ajayamohan, R. (2010).
\newblock Composite structure of monsoon low pressure systems and its relation
  to indian rainfall.
\newblock {\em Journal of Climate}, 23(16):4285--4305.

\bibitem[Krishnamurti et~al., 1977]{krishnamurti1977downstream}
Krishnamurti, T., Molinari, J., Pan, H.-l., and Wong, V. (1977).
\newblock Downstream amplification and formation of monsoon disturbances.
\newblock {\em Monthly Weather Review}, 105(10):1281--1297.

\bibitem[Krishnamurti et~al., 1981]{krishnamurti1981onset}
Krishnamurti, T.~N., Ardanuy, P., Ramanathan, Y., and Pasch, R. (1981).
\newblock On the onset vortex of the summer monsoon.
\newblock {\em Monthly Weather Review}, 109(2):344--363.

\bibitem[Kumar et~al., 2008]{kumar2008analysis}
Kumar, A., Dudhia, J., Rotunno, R., Niyogi, D., and Mohanty, U. (2008).
\newblock Analysis of the 26 july 2005 heavy rain event over mumbai, india
  using the weather research and forecasting (wrf) model.
\newblock {\em Quarterly Journal of the Royal Meteorological Society},
  134(636):1897--1910.

\bibitem[Kumar and Bhat, 2017]{kumar2017vertical}
Kumar, S. and Bhat, G. (2017).
\newblock Vertical structure of orographic precipitating clouds observed over
  south asia during summer monsoon season.
\newblock {\em Journal of Earth System Science}, 126(8):1--12.

\bibitem[Kumar et~al., 2014]{kumar2014role}
Kumar, S., Hazra, A., and Goswami, B. (2014).
\newblock Role of interaction between dynamics, thermodynamics and cloud
  microphysics on summer monsoon precipitating clouds over the myanmar coast
  and the western ghats.
\newblock {\em Climate dynamics}, 43(3-4):911--924.

\bibitem[Kushwaha et~al., 2021]{ksn}
Kushwaha, P., Sukhatme, J., and Nanjundiah, R. (2021).
\newblock A global tropical survey of midtropospheric cyclones.
\newblock {\em Monthly Weather Review}, 149(8):2737 -- 2753.

\bibitem[Lee et~al., 2013]{lee2013real}
Lee, J.-Y., Wang, B., Wheeler, M.~C., Fu, X., Waliser, D.~E., and Kang, I.-S.
  (2013).
\newblock Real-time multivariate indices for the boreal summer intraseasonal
  oscillation over the asian summer monsoon region.
\newblock {\em Climate Dynamics}, 40(1-2):493--509.

\bibitem[Liebmann and Smith, 1996]{liebmann1996description}
Liebmann, B. and Smith, C.~A. (1996).
\newblock Description of a complete (interpolated) outgoing longwave radiation
  dataset.
\newblock {\em Bulletin of the American Meteorological Society},
  77(6):1275--1277.

\bibitem[Ling et~al., 2016]{ling2016impact}
Ling, Z., Wang, Y., and Wang, G. (2016).
\newblock Impact of intraseasonal oscillations on the activity of tropical
  cyclones in summer over the south china sea. part i: local tropical cyclones.
\newblock {\em J. Clim.}, 29(2):855--868.

\bibitem[Mahto and Mishra, 2019]{mahto2019does}
Mahto, S.~S. and Mishra, V. (2019).
\newblock Does era-5 outperform other reanalysis products for hydrologic
  applications in india?
\newblock {\em Journal of Geophysical Research: Atmospheres},
  124(16):9423--9441.

\bibitem[Mak, 1975]{mak1975monsoonal}
Mak, M. (1975).
\newblock The monsoonal mid-tropospheric cyclogenesis.
\newblock {\em Journal of the Atmospheric Sciences}, 32(12):2246--2253.

\bibitem[Mak, 1983]{mak1983moist}
Mak, M. (1983).
\newblock A moist baroclinic model for monsoonal mid-tropospheric cyclogenesis.
\newblock {\em Journal of the Atmospheric Sciences}, 40(5):1154--1162.

\bibitem[Mak and Jim~Kao, 1982]{mak1982instability}
Mak, M. and Jim~Kao, C.-Y. (1982).
\newblock An instability study of the onset-vortex of the southwest monsoon,
  1979.
\newblock {\em Tellus}, 34(4):358--368.

\bibitem[Manning and Hart, 2007]{manning2007evolution}
Manning, D.~M. and Hart, R.~E. (2007).
\newblock Evolution of north atlantic era40 tropical cyclone representation.
\newblock {\em Geophysical research letters}, 34(5).

\bibitem[Mapes, 2011]{mapes2011heaviest}
Mapes, B. (2011).
\newblock Heaviest precipitation events, 1998-2007: A near-global survey.
\newblock In {\em The Global Monsoon System: Research and Forecast}, pages
  15--22. World Scientific.

\bibitem[Meera et~al., 2019]{meera2019downstream}
Meera, M., Suhas, E., and Sandeep, S. (2019).
\newblock Downstream and in situ: Two perspectives on the initiation of monsoon
  low-pressure systems over the bay of bengal.
\newblock {\em Geophysical Research Letters}, 46(21):12303--12310.

\bibitem[Miller and Keshavamurty, 1968]{miller1968iioe}
Miller, F. and Keshavamurty, R. (1968).
\newblock {Structure of an Arabian Sea summer monsoon system}.
\newblock {\em International Indian Ocean Experiment, Metero. Monog.}, 1.

\bibitem[Mooley, 1973]{mooley1973}
Mooley, D. (1973).
\newblock Some aspects of indian monsoon depressions and the associated
  rainfall.
\newblock {\em Monthly Weather Review}, 101:271--280.

\bibitem[Mooley and Shukla, 1987]{mooley1987characteristics}
Mooley, D. and Shukla, J. (1987).
\newblock {\em Characteristics of the westward-moving summer monsoon low
  pressure systems over the Indian region and their relationship with the
  monsoon rainfall}.
\newblock University of Maryland, Department of Meteorology, Center for
  Ocean-Land~….

\bibitem[Narayanan and Rao, 1981]{narayanan1981detection}
Narayanan, M. and Rao, B. (1981).
\newblock Detection of monsoon inversion by tiros-n satellite.
\newblock {\em Nature}, 294(5841):546--548.

\bibitem[Nogueira, 2020]{nogueira2020inter}
Nogueira, M. (2020).
\newblock Inter-comparison of era-5, era-interim and gpcp rainfall over the
  last 40 years: Process-based analysis of systematic and random differences.
\newblock {\em Journal of Hydrology}, 583:124632.

\bibitem[Pattanaik and Rajeevan, 2010]{pattanaik2010variability}
Pattanaik, D. and Rajeevan, M. (2010).
\newblock Variability of extreme rainfall events over india during southwest
  monsoon season.
\newblock {\em Meteorological Applications: A journal of forecasting, practical
  applications, training techniques and modelling}, 17(1):88--104.

\bibitem[Patwardhan et~al., 2020]{patwardhan2020synoptic}
Patwardhan, S., Sooraj, K., Varikoden, H., Vishnu, S., Koteswararao, K.,
  Ramarao, M., and Pattanaik, D. (2020).
\newblock Synoptic scale systems.
\newblock In {\em Assessment of Climate Change over the Indian Region}, pages
  143--154. Springer.

\bibitem[Pearce and Mohanthy, 1984]{pearce1984onsets}
Pearce, R. and Mohanthy, U. (1984).
\newblock Onsets of the asian summer monsoon 1979--82.
\newblock {\em Journal of Atmospheric Sciences}, 41(9):1620--1639.

\bibitem[Pope et~al., 2009]{pope2009regimes}
Pope, M., Jakob, C., and Reeder, M.~J. (2009).
\newblock Regimes of the north australian wet season.
\newblock {\em Journal of Climate}, 22(24):6699--6715.

\bibitem[Prakash et~al., 2013]{prakash2013comparison}
Prakash, S., Mahesh, C., and Gairola, R. (2013).
\newblock Comparison of trmm multi-satellite precipitation analysis (tmpa)-3b43
  version 6 and 7 products with rain gauge data from ocean buoys.
\newblock {\em Remote sensing letters}, 4(7):677--685.

\bibitem[Praveen et~al., 2015]{praveen2015relationship}
Praveen, V., Sandeep, S., and Ajayamohan, R. (2015).
\newblock On the relationship between mean monsoon precipitation and low
  pressure systems in climate model simulations.
\newblock {\em Journal of Climate}, 28(13):5305--5324.

\bibitem[Qian et~al., 2019]{qian2019new}
Qian, Y., Hsu, P.-C., and Kazuyoshi, K. (2019).
\newblock New real-time indices for the quasi-biweekly oscillation over the
  asian summer monsoon region.
\newblock {\em Climate Dynamics}, 53(5):2603--2624.

\bibitem[Rajeevan et~al., 2008]{rajeevan2008analysis}
Rajeevan, M., Bhate, J., and Jaswal, A.~K. (2008).
\newblock Analysis of variability and trends of extreme rainfall events over
  india using 104 years of gridded daily rainfall data.
\newblock {\em Geophysical research letters}, 35(18).

\bibitem[Ramage, 1966]{ramage1966summer}
Ramage, C. (1966).
\newblock The summer atmospheric circulation over the arabian sea.
\newblock {\em Journal of Atmospheric Sciences}, 23(2):144--150.

\bibitem[Ramage, 1971]{ramage1971monsoon}
Ramage, C.~S. (1971).
\newblock Monsoon meteorology.
\newblock Technical report.

\bibitem[Ray et~al., 2019]{ray2019recent}
Ray, K., Pandey, P., Pandey, C., Dimri, A., and Kishore, K. (2019).
\newblock On the recent floods in india.
\newblock {\em Current science}, 117(2):204--218.

\bibitem[Rousseeuw, 1987]{rousseeuw1987silhouettes}
Rousseeuw, P.~J. (1987).
\newblock Silhouettes: a graphical aid to the interpretation and validation of
  cluster analysis.
\newblock {\em Journal of computational and applied mathematics}, 20:53--65.

\bibitem[Roxy et~al., 2017]{roxy2017threefold}
Roxy, M.~K., Ghosh, S., Pathak, A., Athulya, R., Mujumdar, M., Murtugudde, R.,
  Terray, P., and Rajeevan, M. (2017).
\newblock A threefold rise in widespread extreme rain events over central
  india.
\newblock {\em Nature communications}, 8(1):1--11.

\bibitem[Satopaa et~al., 2011]{satopaa2011finding}
Satopaa, V., Albrecht, J., Irwin, D., and Raghavan, B. (2011).
\newblock Finding a" kneedle" in a haystack: Detecting knee points in system
  behavior.
\newblock In {\em 2011 31st international conference on distributed computing
  systems workshops}, pages 166--171. IEEE.

\bibitem[Shukla, 1978]{shukla1978}
Shukla, J. (1978).
\newblock Cisk-barotropic-baroclinic instability and the growth of monsoon
  depressions.
\newblock {\em Journal of the Atmospheric Sciences}, 35:495--508.

\bibitem[Shyamala and Bhadram, 2006]{shyamala2006impact}
Shyamala, B. and Bhadram, C. (2006).
\newblock Impact of mesoscale--synoptic scale interactions on the mumbai
  historical rain event during 26--27 july 2005.
\newblock {\em Current Science}, pages 1649--1654.

\bibitem[Sikka, 1980]{sikka1980some}
Sikka, D. (1980).
\newblock Some aspects of the large scale fluctuations of summer monsoon
  rainfall over india in relation to fluctuations in the planetary and regional
  scale circulation parameters.
\newblock {\em Proceedings of the Indian Academy of Sciences-Earth and
  Planetary Sciences}, 89(2):179--195.

\bibitem[Sobel and Horinouchi, 2000]{sobel2000dynamics}
Sobel, A.~H. and Horinouchi, T. (2000).
\newblock On the dynamics of easterly waves, monsoon depressions, and tropical
  depression type disturbances.
\newblock {\em Journal of the Meteorological Society of Japan. Ser. II},
  78(2):167--173.

\bibitem[S{\o}rland and Sorteberg, 2015]{sorland2015dynamic}
S{\o}rland, S.~L. and Sorteberg, A. (2015).
\newblock The dynamic and thermodynamic structure of monsoon low-pressure
  systems during extreme rainfall events.
\newblock {\em Tellus A: Dynamic Meteorology and Oceanography}, 67(1):27039.

\bibitem[Syakur et~al., 2018]{syakur2018integration}
Syakur, M., Khotimah, B., Rochman, E., and Satoto, B. (2018).
\newblock Integration k-means clustering method and elbow method for
  identification of the best customer profile cluster.
\newblock In {\em IOP Conference Series: Materials Science and Engineering},
  volume 336, page 012017. IOP Publishing.

\bibitem[Tawde and Singh, 2015]{tawde2015investigation}
Tawde, S.~A. and Singh, C. (2015).
\newblock Investigation of orographic features influencing spatial distribution
  of rainfall over the western ghats of india using satellite data.
\newblock {\em International Journal of Climatology}, 35(9):2280--2293.

\bibitem[Varikoden et~al., 2019]{varikoden2019contrasting}
Varikoden, H., Revadekar, J., Kuttippurath, J., and Babu, C. (2019).
\newblock Contrasting trends in southwest monsoon rainfall over the western
  ghats region of india.
\newblock {\em Climate Dynamics}, 52(7):4557--4566.

\bibitem[Vinnarasi and Dhanya, 2016]{vinnarasi2016changing}
Vinnarasi, R. and Dhanya, C. (2016).
\newblock Changing characteristics of extreme wet and dry spells of indian
  monsoon rainfall.
\newblock {\em Journal of Geophysical Research: Atmospheres},
  121(5):2146--2160.

\bibitem[Vuruputur et~al., 2018]{vuruputur2018tropical}
Vuruputur, V., Sukhatme, J., Murtugudde, R., and Roca, R. (2018).
\newblock {\em Tropical Extremes: Natural Variability and Trends}.
\newblock Elsevier.

\bibitem[Wernli and Schwierz, 2006]{wernli2006surface}
Wernli, H. and Schwierz, C. (2006).
\newblock Surface cyclones in the era-40 dataset (1958--2001). part i: Novel
  identification method and global climatology.
\newblock {\em Journal of the atmospheric sciences}, 63(10):2486--2507.

\bibitem[Wheeler and Hendon, 2004]{wheeler2004all}
Wheeler, M.~C. and Hendon, H.~H. (2004).
\newblock An all-season real-time multivariate mjo index: Development of an
  index for monitoring and prediction.
\newblock {\em Monthly weather review}, 132(8):1917--1932.

\bibitem[Wood and Ritchie, 2014]{wood201440}
Wood, K.~M. and Ritchie, E.~A. (2014).
\newblock A 40-year climatology of extratropical transition in the eastern
  north pacific.
\newblock {\em Journal of climate}, 27(15):5999--6015.

\bibitem[Wu et~al., 1999]{wu1999numerical}
Wu, Y., Raman, S., and Mohanty, U. (1999).
\newblock Numerical investigation of the somali jet interaction with the
  western ghat mountains.
\newblock {\em pure and applied geophysics}, 154(2):365--396.

\bibitem[Yeasmin et~al., 2021]{yeasmin2021detection}
Yeasmin, A., Chand, S., Turville, C., and Sultanova, N. (2021).
\newblock Detection and verification of tropical cyclones and depressions over
  the south pacific ocean basin using era-5 reanalysis dataset.
\newblock {\em International Journal of Climatology}.

\bibitem[Yihui and Chan, 2005]{yihui2005east}
Yihui, D. and Chan, J.~C. (2005).
\newblock The east asian summer monsoon: an overview.
\newblock {\em Meteorology and Atmospheric Physics}, 89(1):117--142.

\bibitem[Yoon and Chen, 2005]{yoon2005water}
Yoon, J.-H. and Chen, T.-C. (2005).
\newblock Water vapor budget of the indian monsoon depression.
\newblock {\em Tellus A: Dynamic Meteorology and Oceanography}, 57(5):770--782.

\bibitem[Zhang and Smith, 2018]{zhang2018numerical}
Zhang, G. and Smith, R.~B. (2018).
\newblock Numerical study of physical processes controlling summer
  precipitation over the western ghats region.
\newblock {\em Journal of Climate}, 31(8):3099--3115.

\end{thebibliography}

\end{document}


\title{A Global Survey of Mid-Tropospheric Cyclones and Case Studies from the Arabian Sea}

\authors{Pradeep Kushwaha, pkushwaha9999@gmail.com,
        \correspondingauthor{Pradeep Kushwaha, Centre for Atmospheric \& Oceanic Sciences, Indian Institute of Science, Bangalore, India.}}
        \email{pkushwaha9999@gmail.com}
        \affiliation{Centre for Atmospheric \& Oceanic Sciences, Indian Institute of Science, Bangalore, India.}

        \extraauthor{Jai Sukhatme,  jai.goog@gmail.com}
        \extraaffil{Centre for Atmospheric \& Oceanic Sciences and Divecha Centre for Climate Change, Indian Institute of Science, Bangalore, India.}

        \extraauthor{Ravi Nanjundiah, ravisn@gmail.com}
        \extraaffil{Centre for Atmospheric \& Oceanic Sciences, Indian Institute of Science, Bangalore, India and Indian Institute of Tropical Meteorology, Pune, India.}
\begin{center}
{\Large {\bf Supplementary Material}}
\end{center}

\section{MTC, monsoon depression and tropical cyclone profiles}

The list of cyclonic centers used to make composites from which we derive MTC, monsoon depression (MD) and tropical cyclone (TC) vorticity profiles are listed in Tables~\ref{Table:TS2}, \ref{Table:TS3} and \ref{Table:TS4}, respectively. 
The IMD best track data can be found at: \url{http://www.rsmcnewdelhi.imd.gov.in/index.php?option=com_content&view=article&id=48&Itemid=194&lang=en} and the IBTRACS data is located at: \url{https://www.ncei.noaa.gov/data/international-best-track-archive-for-climate-stewardship-ibtracs/v04r00/access/csv/}.

\section{Sensitivity of Number of MTCs and LTCs}
\subsection{Layer combinations and thickness}
To understand the effect of different choices of layers we have used data of 882 non-topographic cyclonic system detected during the 35 MTCs over $5^{\circ}$N-$25^{\circ}$N and $50^{\circ}$E-$95^{\circ}$E. Figure~\ref{fig:FIG_S1} shows pdfs of $\delta \xi_{p}$ of these systems for various combination of layers:
\begin{enumerate}
\item Layer combination 1: Lower layer 1000-850 hPa and middle layer 650-500 hPa (LC0: black curve): This combination is used in all results presented in the study.
    
\item Layer combination 2: Lower layer 1000-850 hPa and middle layer 850-500 hPa (LC1: blue curve); broad middle layer.
    
\item  Layer combination-3: Lower layer 1000-650 hPa and middle layer 650-500 hPa (LC2: red curve); broad lower layer. 
    
\item  Layer combination-4: Lower layer 900 hPa and middle layer 600 hPa (LC3: green curve). 
\end{enumerate}
As discussed in the Methods section, the aim is to filter systems as per the location ($P_{\xi}$) and strength ($\delta P_{\xi}$) of vorticity maxima. As is seen in Figure~\ref{fig:FIG_S1}, the the choice of layers does not affect the pdf to a great extent. 
Examining LC3, this probability distribution is quite similar to layer combination LC0. However, this choice has the highest variance of $\delta P_{\xi}$ because no layer mean is performed; hence differences of middle and lower layer vorticity turn out to be substantial (Table~\ref{Table:TS1}). 
Further, a layer combination which is not equal in thickness, i.e., relatively thick middle or lower layer affects the number of MTCs. This is mainly because of the shrinking of pdfs towards the origin (Figure~\ref{fig:FIG_S1}). Moreover the reduction in the number of MTCs is large compare to LTCs because MTCs threshold $\delta \xi_{p}>1.5 \times 10^{-5}$ $s^{-1}$ is near the outer region of the tail of pdf and LTCs threshold $\delta \xi_{p}<0.5 \times 10^{-5}$ $s^{-1}$ is near the origin (for numbers, see Table~\ref{Table:TS1}). Thus, it is inappropriate to use a single level or skewed layer thicknesses and the moderate choice of 150 hPa thickness layers, i.e., combination LC0, is appropriate.

\subsection{Moisture and relative vorticity thresholds}
Table~\ref{Table:TS1} also shows the number of MTCs and LTCs corresponding to various values of moisture ($Q_{m}$) and vorticity ($\xi_{m}$) thresholds. As expected, the number of MTCs and LTCs reduces with higher thresholds. Also, the number of LTCs is relatively less sensitive to moisture thresholds than MTCs at lower moisture thresholds. This indicates that, in general, LTCs are more moist in nature as compared to MTCs. 

\subsection{Relaxation of $P_{\xi}$ threshold}
We have relaxed the constraint on the level of vorticity maxima ($P_{\xi}$) to delineate MTC and LTC phases during tracking. 
Figure~\ref{fig:FIG_S2} shows overall, MTC and LTC center density and their fraction in the presence of $P_{\xi}$ threshold and without it, respectively. As expected, the qualitative nature of results remains the same with a slight shrinking of MTC and LTC activity regions. 

\section{Global tropical MTC and LTC composite}

To further demonstrate the robustness of MTC and LTC structures, we show results from a composite from entire tropics in the summer and winter seasons. These are shown in Figure~\ref{fig:FIG_S3} and Figure~\ref{fig:FIG_S4}, respectively. 

\begin{table}[]
      
\centering
\begin{tabular}{|l|l|l|l|l|l|l|l|l|l|l|l|l|l|}
\cline{1-4} \cline{6-9} \cline{11-14}
SN. NO & DATES    & LAT  & LON  &  & SN. NO & DATES    & LAT  & LON  &  & SN. NO & DATES    & LAT  & LON  \\ \cline{1-4} \cline{6-9} \cline{11-14} 
1      & 19880819 & 19.5 & 69   &  & 23     & 19970707 & 21   & 67.5 &  & 45     & 20050630 & 21   & 73.5 \\ \cline{1-4} \cline{6-9} \cline{11-14} 
2      & 19890824 & 18   & 66   &  & 24     & 20000628 & 21   & 67.5 &  & 46     & 20050701 & 21   & 70.5 \\ \cline{1-4} \cline{6-9} \cline{11-14} 
3      & 19900626 & 19.5 & 72   &  & 25     & 20000629 & 16.5 & 72   &  & 47     & 20060629 & 18   & 67.5 \\ \cline{1-4} \cline{6-9} \cline{11-14} 
4      & 19900627 & 19.5 & 70.5 &  & 26     & 20000630 & 16.5 & 72   &  & 48     & 20060630 & 21   & 70.5 \\ \cline{1-4} \cline{6-9} \cline{11-14} 
5      & 19920718 & 21   & 66   &  & 27     & 20000704 & 21   & 66   &  & 49     & 20060701 & 21   & 70.5 \\ \cline{1-4} \cline{6-9} \cline{11-14} 
6      & 19920718 & 21   & 72   &  & 28     & 20020920 & 19.5 & 70.5 &  & 50     & 20060703 & 19.5 & 72   \\ \cline{1-4} \cline{6-9} \cline{11-14} 
7      & 19920719 & 21   & 67.5 &  & 29     & 20020922 & 18   & 67.5 &  & 51     & 20070804 & 21   & 69   \\ \cline{1-4} \cline{6-9} \cline{11-14} 
8      & 19920720 & 21   & 66   &  & 30     & 20020924 & 18   & 66   &  & 52     & 20070807 & 21   & 72   \\ \cline{1-4} \cline{6-9} \cline{11-14} 
9      & 19930826 & 19.5 & 69   &  & 31     & 20030701 & 16.5 & 72   &  & 53     & 20080728 & 21   & 70.5 \\ \cline{1-4} \cline{6-9} \cline{11-14} 
10     & 19930827 & 19.5 & 67.5 &  & 32     & 20030702 & 16.5 & 70.5 &  & 54     & 20080729 & 19.5 & 67.5 \\ \cline{1-4} \cline{6-9} \cline{11-14} 
11     & 19930828 & 18   & 66   &  & 33     & 20030712 & 21   & 67.5 &  & 55     & 20080729 & 21   & 73.5 \\ \cline{1-4} \cline{6-9} \cline{11-14} 
12     & 19940827 & 21   & 75   &  & 34     & 20030717 & 21   & 66   &  & 56     & 20080730 & 19.5 & 69   \\ \cline{1-4} \cline{6-9} \cline{11-14} 
13     & 19940828 & 21   & 69   &  & 35     & 20030914 & 15   & 67.5 &  & 57     & 20080912 & 18   & 72   \\ \cline{1-4} \cline{6-9} \cline{11-14} 
14     & 19940829 & 21   & 67.5 &  & 36     & 20040704 & 19.5 & 70.5 &  & 58     & 20080913 & 21   & 73.5 \\ \cline{1-4} \cline{6-9} \cline{11-14} 
15     & 19950705 & 18   & 70.5 &  & 37     & 20040705 & 18   & 67.5 &  & 59     & 20080914 & 18   & 67.5 \\ \cline{1-4} \cline{6-9} \cline{11-14} 
16     & 19950711 & 19.5 & 67.5 &  & 38     & 20050624 & 21   & 69   &  & 60     & 20080915 & 21   & 70.5 \\ \cline{1-4} \cline{6-9} \cline{11-14} 
17     & 19950712 & 18   & 66   &  & 39     & 20050625 & 21   & 70.5 &  & 61     & 20080916 & 21   & 70.5 \\ \cline{1-4} \cline{6-9} \cline{11-14} 
18     & 19950906 & 21   & 69   &  & 40     & 20050626 & 21   & 69   &  &        &          &      &      \\ \cline{1-4} \cline{6-9} \cline{11-14} 
19     & 19950907 & 19.5 & 69   &  & 41     & 20050627 & 21   & 70.5 &  &        &          &      &      \\ \cline{1-4} \cline{6-9} \cline{11-14} 
20     & 19960722 & 19.5 & 69   &  & 42     & 20050629 & 19.5 & 67.5 &  &        &          &      &      \\ \cline{1-4} \cline{6-9} \cline{11-14} 
21     & 19970704 & 18   & 67.5 &  & 43     & 20050629 & 21   & 72   &  &        &          &      &      \\ \cline{1-4} \cline{6-9} \cline{11-14} 
22     & 19970705 & 18   & 67.5 &  & 44     & 20050630 & 21   & 67.5 &  &        &          &      &      \\ \cline{1-4} \cline{6-9} \cline{11-14} 
\end{tabular}
\caption {61 Non-topographic strong cyclonic centers ($\xi_{m}>3 \times 10^{-5} s^{-1}$, $\xi_{l}$ $>0$) detected during IMD dates of middle level circulation at 12:00 UTC \citep{choudhury2018phenomenological} over north Arabian sea $5^{\circ}$N-$22^{\circ}$N, $65^{\circ}$E-$75^{\circ}$E at 600 hPa geopotential height. }
\label{Table:TS2}

\end{table}

\begin{table}[]
\centering

\begin{tabular}{|l|l|l|l|l|l|l|l|l|l|l|l|l|}
\cline{1-6} \cline{8-13}
SN. NO & DATES    & LAT  & LON  & TYPE & USED &  & SN. NO & DATES      & LAT    & LON    & TYPE   & USED \\ \cline{1-6} \cline{8-13} 
1      & 20140721 & 22.5  & 85.5 & D    & Y    &  & 23     & 20150917   & 19.5   & 81.0   & D      & Y    \\ \cline{1-6} \cline{8-13} 
2      & 20140722 & 22.5 & 81.0 & D    & Y    &  & 24     & 20150918   & 22.5   & 75.0   & D      & Y    \\ \cline{1-6} \cline{8-13} 
3      & 20140723 & 22.5 & 79.5 & D    & Y    &  & 25     & 20150919   & 22.5   & 72.0   & D      & Y    \\ \cline{1-6} \cline{8-13} 
4      & 20140803 & 22.5 & 88.5 & DD   &      &  & 26     & 20160706   & 24     & 81.0   & D      & Y    \\ \cline{1-6} \cline{8-13} 
5      & 20140804 & 22.5 & 87.0 & DD   & Y    &  & 27     & 20160707   & 22.5   & 81.0   & D      &      \\ \cline{1-6} \cline{8-13} 
6      & 20140805 & 24   & 82.5 & DD   & Y    &  & 29     & 20160810   & 24.0   & 88.5   & DD     & Y    \\ \cline{1-6} \cline{8-13} 
7      & 20140806 & 25.5 & 79.5 & D    & Y    &  & 30     & 20160811   & 24     & 85.5   & D      & Y    \\ \cline{1-6} \cline{8-13} 
8      & 20150620 & 18.0 & 84   & D    & Y    &  & 31     & 20160812   & 24     & 85.5   & D      &      \\ \cline{1-6} \cline{8-13} 
9      & 20150621 & 21.0 & 84   & D    & Y    &  & 32     & 20160816   & 21.0   & 91.5   & D      &      \\ \cline{1-6} \cline{8-13} 
10     & 20150622 & 22.5 & 85.5 & D    & Y    &  & 33     & 20160817   & 22.5   & 87.0   & D      & Y    \\ \cline{1-6} \cline{8-13} 
11     & 20150710 & 24.0 & 82.5 & D    & Y    &  & 34     & 20160818   & 24.0   & 84.0   & D      & Y    \\ \cline{1-6} \cline{8-13} 
12     & 20150711 & 25.5 & 79.5 & D    & Y    &  & 35     & 20160819   & 25.5   & 79.5   & D      & Y    \\ \cline{1-6} \cline{8-13} 
13     & 20150712 & 27.0 & 79.5 & D    &      &  & 36     & 20160820   & 25.5   & 76.5   & D      & Y    \\ \cline{1-6} \cline{8-13} 
14     & 20150726 & 22.5 & 90   & D    & Y    &  & 37     & 20170611   & 22.5   & 90     & D      & Y    \\ \cline{1-6} \cline{8-13} 
15     & 20150727 & 22.5 & 90   & D    & Y    &  & 38     & 20170612   & 24.0   & 91.5   & D      & Y    \\ \cline{1-6} \cline{8-13} 
16     & 20150728 & 21.0 & 91.5 & D    & Y    &  & 39     & 20170718   & 19.5   & 84.5   & D      & Y    \\ \cline{1-6} \cline{8-13} 
17     & 20150729 & 21.0 & 91.5 & DD   & Y    &  & 40     & 20170719   & 21.0   & 84.0   & D      &      \\ \cline{1-6} \cline{8-13} 
18     & 20150730 & 22.5 & 91.5 & CS   & Y    &  & 41     & 20170726   & 24.0   & 82.5   & D      & Y    \\ \cline{1-6} \cline{8-13} 
19     & 20150731 & 24.0 & 90.0 & D    & Y    &  & 42     & 20170727   & 24.0   & 79.5   & D      & Y    \\ \cline{1-6} \cline{8-13} 
20     & 20150801 & 22.5 & 88.5 & D    & Y    &  & 43     & 20180610   & 24.0   & 91.5   & D      & Y    \\ \cline{1-6} \cline{8-13} 
21     & 20150802 & 22.5 & 88.5 & D    & Y    &  & 44     & 20180611   & 24.0   & 91.5   & D      &      \\ \cline{1-6} \cline{8-13} 
22     & 20150916 & 20.5 & 82.5 & D    & Y    &  & TOTAL  & {\ul USED} & {\ul } & {\ul } & {\ul } & 36   \\ \cline{1-6} \cline{8-13} 
\end{tabular}
\caption {36 Monsoon Depressions centers detected at 12:00 UTC on 600 hPa over Indian region $5^{\circ}$N-$30^{\circ}$N, $70^{\circ}$E-$95^{\circ}$E in MERRA-2 data and matched with IMD Best Track Data from 2014--2018 (selected only if $\xi_{m}>6 \times 10^{-5}$ s$^{-1}$). "Y" denotes that we have used the system in our calculation; the date of detection is mentioned in the format "YYYYMMDD" (year-month-day) and corresponding position latitude (LAT) and longitude (LON) coordinate; for example detected center on 20140721 is 21 July 2014 and is found at $22.5^{\circ}$N, latitude and $85.5^{\circ}$E, longitude in MERRA-2 data at 1.5 degree resolution. The intensity type of each center is corresponds to IMD definition of intensity: D --- Depression, DD --- Deep Depression, CS --- Cyclonic Storm.}
\label{Table:TS3}

\end{table}
\begin{table}[]

\centering
\scalebox{0.85}{

\begin{tabular}{|l|l|l|l|l|l|l|l|l|lllll}
\cline{1-4} \cline{6-9} \cline{11-14}
SR. NO & DATE     & LAT  & LON   &  & SR.NO & DATE     & LAT  & LON   & \multicolumn{1}{l|}{} & \multicolumn{1}{l|}{SR.NO} & \multicolumn{1}{l|}{DATE}     & \multicolumn{1}{l|}{LAT}  & \multicolumn{1}{l|}{LON}   \\ \cline{1-4} \cline{6-9} \cline{11-14} 
1      & 20140703 & 10.5 & 144   &  & 34    & 20150805 & 19.5 & 132   & \multicolumn{1}{l|}{} & \multicolumn{1}{l|}{67}    & \multicolumn{1}{l|}{20180923} & \multicolumn{1}{l|}{18}   & \multicolumn{1}{l|}{133.5} \\ \cline{1-4} \cline{6-9} \cline{11-14} 
2      & 20140704 & 13.5 & 141   &  & 35    & 20150815 & 13.5 & 145.5 & \multicolumn{1}{l|}{} & \multicolumn{1}{l|}{68}    & \multicolumn{1}{l|}{20180924} & \multicolumn{1}{l|}{19.5} & \multicolumn{1}{l|}{129}   \\ \cline{1-4} \cline{6-9} \cline{11-14} 
3      & 20140705 & 16.5 & 136.5 &  & 36    & 20150815 & 15   & 162   & \multicolumn{1}{l|}{} & \multicolumn{1}{l|}{69}    & \multicolumn{1}{l|}{20180925} & \multicolumn{1}{l|}{19.5} & \multicolumn{1}{l|}{129}   \\ \cline{1-4} \cline{6-9} \cline{11-14} 
4      & 20140706 & 19.5 & 130.5 &  & 37    & 20150816 & 15   & 144   & \multicolumn{1}{l|}{} & \multicolumn{1}{l|}{70}    & \multicolumn{1}{l|}{20180930} & \multicolumn{1}{l|}{15}   & \multicolumn{1}{l|}{138}   \\ \cline{1-4} \cline{6-9} \cline{11-14} 
5      & 20140715 & 13.5 & 124.5 &  & 38    & 20150816 & 15   & 160.5 & \multicolumn{1}{l|}{} & \multicolumn{1}{l|}{71}    & \multicolumn{1}{l|}{20190805} & \multicolumn{1}{l|}{18}   & \multicolumn{1}{l|}{129}   \\ \cline{1-4} \cline{6-9} \cline{11-14} 
6      & 20140720 & 15   & 129   &  & 39    & 20150817 & 15   & 157.5 & \multicolumn{1}{l|}{} & \multicolumn{1}{l|}{72}    & \multicolumn{1}{l|}{20190806} & \multicolumn{1}{l|}{19.5} & \multicolumn{1}{l|}{129}   \\ \cline{1-4} \cline{6-9} \cline{11-14} 
7      & 20140721 & 19.5 & 124.5 &  & 40    & 20150817 & 18   & 139.5 & \multicolumn{1}{l|}{} & \multicolumn{1}{l|}{73}    & \multicolumn{1}{l|}{20190806} & \multicolumn{1}{l|}{19.5} & \multicolumn{1}{l|}{142.5} \\ \cline{1-4} \cline{6-9} \cline{11-14} 
8      & 20140729 & 19.5 & 130.5 &  & 41    & 20150818 & 16.5 & 154.5 & \multicolumn{1}{l|}{} & \multicolumn{1}{l|}{74}    & \multicolumn{1}{l|}{20190822} & \multicolumn{1}{l|}{18}   & \multicolumn{1}{l|}{127.5} \\ \cline{1-4} \cline{6-9} \cline{11-14} 
9      & 20140801 & 15   & 138   &  & 42    & 20150818 & 18   & 132   & \multicolumn{1}{l|}{} & \multicolumn{1}{l|}{75}    & \multicolumn{1}{l|}{20190823} & \multicolumn{1}{l|}{19.5} & \multicolumn{1}{l|}{124.5} \\ \cline{1-4} \cline{6-9} \cline{11-14} 
10     & 20140802 & 15   & 135   &  & 43    & 20150819 & 19.5 & 127.5 & \multicolumn{1}{l|}{} & \multicolumn{1}{l|}{76}    & \multicolumn{1}{l|}{20190929} & \multicolumn{1}{l|}{19.5} & \multicolumn{1}{l|}{124.5} \\ \cline{1-4} \cline{6-9} \cline{11-14} 
11     & 20140803 & 16.5 & 133.5 &  & 44    & 20150819 & 19.5 & 151.5 &                       &                            &                               &                           &                            \\ \cline{1-4} \cline{6-9}
12     & 20140804 & 18   & 130.5 &  & 45    & 20150820 & 19.5 & 124.5 &                       &                            &                               &                           &                            \\ \cline{1-4} \cline{6-9}
13     & 20140913 & 15   & 126   &  & 46    & 20150821 & 19.5 & 123   &                       &                            &                               &                           &                            \\ \cline{1-4} \cline{6-9}
14     & 20140914 & 18   & 121.5 &  & 47    & 20150923 & 18   & 135   &                       &                            &                               &                           &                            \\ \cline{1-4} \cline{6-9}
15     & 20140918 & 16.5 & 126   &  & 48    & 20150925 & 19.5 & 132   &                       &                            &                               &                           &                            \\ \cline{1-4} \cline{6-9}
16     & 20140919 & 18   & 120   &  & 49    & 20160804 & 19.5 & 148.5 &                       &                            &                               &                           &                            \\ \cline{1-4} \cline{6-9}
17     & 20140924 & 19.5 & 148.5 &  & 50    & 20160913 & 19.5 & 123   &                       &                            &                               &                           &                            \\ \cline{1-4} \cline{6-9}
18     & 20140930 & 16.5 & 145.5 &  & 51    & 20160924 & 19.5 & 133.5 &                       &                            &                               &                           &                            \\ \cline{1-4} \cline{6-9}
19     & 20150703 & 10.5 & 148.5 &  & 52    & 20160925 & 19.5 & 129   &                       &                            &                               &                           &                            \\ \cline{1-4} \cline{6-9}
20     & 20150704 & 13.5 & 147   &  & 53    & 20180607 & 18   & 129   &                       &                            &                               &                           &                            \\ \cline{1-4} \cline{6-9}
21     & 20150705 & 15   & 144   &  & 54    & 20180608 & 19.5 & 127.5 &                       &                            &                               &                           &                            \\ \cline{1-4} \cline{6-9}
22     & 20150705 & 18   & 120   &  & 55    & 20180705 & 15   & 142.5 &                       &                            &                               &                           &                            \\ \cline{1-4} \cline{6-9}
23     & 20150706 & 18   & 139.5 &  & 56    & 20180706 & 16.5 & 141   &                       &                            &                               &                           &                            \\ \cline{1-4} \cline{6-9}
24     & 20150707 & 15   & 153   &  & 57    & 20180707 & 18   & 141   &                       &                            &                               &                           &                            \\ \cline{1-4} \cline{6-9}
25     & 20150707 & 18   & 135   &  & 58    & 20180816 & 18   & 141   &                       &                            &                               &                           &                            \\ \cline{1-4} \cline{6-9}
26     & 20150708 & 16.5 & 148.5 &  & 59    & 20180819 & 16.5 & 150   &                       &                            &                               &                           &                            \\ \cline{1-4} \cline{6-9}
27     & 20150709 & 18   & 145.5 &  & 60    & 20180820 & 18   & 147   &                       &                            &                               &                           &                            \\ \cline{1-4} \cline{6-9}
28     & 20150710 & 18   & 141   &  & 61    & 20180831 & 18   & 141   &                       &                            &                               &                           &                            \\ \cline{1-4} \cline{6-9}
29     & 20150711 & 18   & 139.5 &  & 62    & 20180910 & 13.5 & 144   &                       &                            &                               &                           &                            \\ \cline{1-4} \cline{6-9}
30     & 20150712 & 18   & 138   &  & 63    & 20180912 & 13.5 & 133.5 &                       &                            &                               &                           &                            \\ \cline{1-4} \cline{6-9}
31     & 20150802 & 15   & 145.5 &  & 64    & 20180913 & 15   & 129   &                       &                            &                               &                           &                            \\ \cline{1-4} \cline{6-9}
32     & 20150803 & 18   & 141   &  & 65    & 20180914 & 18   & 124.5 &                       &                            &                               &                           &                            \\ \cline{1-4} \cline{6-9}
33     & 20150804 & 19.5 & 136.5 &  & 66    & 20180921 & 15   & 142.5 &                       &                            &                               &                           &                            \\ \cline{1-4} \cline{6-9}
\end{tabular}
}
\caption {76 Strong tropical cyclones centers ($\xi_{m}>9 \times 10^{-5}$ s$^{-1}$) detected at 12:00 UTC on 600 hPa over the West Pacific ($5^{\circ}$N-$20^{\circ}$N, $120^{\circ}$E-$180^{\circ}$E) and matched with IBTRACS data \citep{knapp2010international}.}

\label{Table:TS4}
\end{table}


\begin{table}[]
\centering

\centering
\begin{tabular}{|l|l|l|l|l|l|l|l|l|l|l|l|l|}
\cline{1-5} \cline{7-13}
\textbf{}         & \textbf{}    & \textbf{}    & \textbf{}    & \textbf{}    &  & \textbf{$Q_{m}$} & \textbf{MTC} & \textbf{LTC} & \textbf{} & \textbf{$\xi_{m}$} & \textbf{MTC} & \textbf{LTC} \\ \cline{1-5} \cline{7-13} 
\textbf{}         & \textbf{LO0} & \textbf{LO1} & \textbf{LO2} & \textbf{LO3} &  & 1.0        & 145          & 251          & \textbf{} & 1.0           & 121          & 290          \\ \cline{1-5} \cline{7-13} 
\textbf{MEAN}     & 0.36         & 0.41         & 0.09         & 0.57         &  & 2.0        & 121          & 246          &           & 2.0           & 120          & 203          \\ \cline{1-5} \cline{7-13} 
\textbf{VARIANCE} & 1.95         & 0.85         & 1.17         & 2.41         &  & 3.0        & 106          & 233          &           & 3.0           & 77           & 126          \\ \cline{1-5} \cline{7-13} 
\textbf{MEDIAN}   & 0.38         & 0.36         & 0.093        & 0.55         &  & 4.0        & 98           & 226          &           & 4.0           & 29           & 62           \\ \cline{1-5} \cline{7-13} 
\textbf{MTC}      & \textbf{121} & \textbf{52}  & \textbf{66}  & \textbf{157} &  & 5.0        & 96           & 223          &           & 5.0           & 6            & 35           \\ \cline{1-5} \cline{7-13} 
\textbf{LTC}      & \textbf{246} & \textbf{220} & \textbf{282} & \textbf{238} &  & 6.0        & 84           & 218          &           & 6.0           & 2            & 21           \\ \cline{1-5} \cline{7-13} 
\end{tabular}

\caption{Mean, variance, median and probability distributions of $\delta P_{\xi}$ of 882 detected systems over  $5^{\circ}$N-$25^{\circ}$N and $50^{\circ}$E-$95^{\circ}E$ (as shown in Figure~\ref{fig:FIG_S1}) and corresponding number of MTCs and LTCs for various combinations of layer thickness (left portion of the table); the number of MTCs and LTCs are highlighted in bold. The sensitivity of number of MTCs and LTCs for various moisture ($Q_{m}$) and relative vorticity ($\xi_{m}$) thresholds in in the right portion of the Table. }
\label{Table:TS1}

\end{table}

\begin{figure*}
\centering
\includegraphics[trim=0 0 0 0, clip,height = 1\textwidth,width = 1\textwidth, angle =0, clip]{FIGURES/SUPPLYMETRY_FIGURE/FIG_S0.pdf}
\caption{Sensitivity of the distribution of differential vorticity ($\delta\xi_{p}$) to different choices of layers combinations. LC0: lower layer 1000-850 hPa, middle layer, 650-500 hPa; LC1: lower layer 1000-850 hPa, middle layer, 850-500 hPa; LC2: lower layer 1000-650 hPa, middle layer, 650-500 hPa and LC3: lower layer, 900 hPa, middle layer, 600 hPa.}
\label{fig:FIG_S1}
\end{figure*}

\begin{figure*}
\centering
\includegraphics[trim=0 0 0 0, clip,height = 0.7\textwidth,width = 1\textwidth, angle =0, clip]{FIGURES/SUPPLYMETRY_FIGURE/FIG_S1.pdf}
\caption{Overall-center, LTC, MTC center density and MTC to LTC fraction per 4-degree per summer from June to September 2000-2019; left panel (a, c, e, g) without $P_{\xi}$ constraints; Right panel (b, d, f, h) with all constraints, respectively.}
\label{fig:FIG_S2}
\end{figure*}

\begin{figure*}
\centering
\includegraphics[trim=0 0 0 0, clip,height = 1.1\textwidth,width = 0.8\textwidth, angle =0, clip]{FIGURES/SUPPLYMETRY_FIGURE/FIG_SR3.pdf}
\caption{(a) Vertical composite cross-section from 6408 MTCs cyclonic centers (b) Vertical composite cross-section from 6427 LTC cyclonic centers; (c, (d) and (e), (f) are horizontal cross-section composites of MTCs composites at 600 hPa and 970 hPa (left) and LTCs (right) at the same levels. The composites are from 2000-2015, JJAS over the region $0-30^{\circ}$N and $10^{\circ}$E-$350^{\circ}$E}
\label{fig:FIG_S3}
\end{figure*}

\begin{figure*}
\centering
\includegraphics[trim=0 0 0 0, clip,height = 1.1\textwidth,width = 0.8\textwidth, angle =0, clip]{FIGURES/SUPPLYMETRY_FIGURE/FIG_SR2.pdf}
\caption{(a) Vertical composite cross-section from 3923 MTC cyclonic centers (b) Vertical composite cross-section from 5384 LTC cyclonic centers; (c), (d) and (e), (f) are horizontal cross-section composites of MTCs composites at 600 hPa and 970 hPa (left) and LTCs (right) at the same levels. The composites are from 2000-2015, DJFM over the region $0-30^{\circ}$N and $10^{\circ}$E-$350^{\circ}$E}
\label{fig:FIG_S4}
\end{figure*}


\bibliographystyle{ametsoc2014}
\bibliography{ref}